\documentclass[useAMS,usenatbib]{mn2e}
\usepackage{graphicx}
\usepackage{gensymb}
\usepackage{amsmath}
\usepackage{amssymb}
\usepackage{url}
\usepackage{longtable}
\usepackage{aas_macros}
\usepackage{deluxetable}

\newcommand\ion[2]{#1$\,${\small{#2}}\relax}

\title[SN\,Ia Nebular Phase Spectra]{Nebular-Phase Spectra of Nearby Type Ia Supernovae}
 
\author[Graham et al.]{M.~L. Graham$^{1,2}$\thanks{E-mail: mlg3k@uw.edu}, 
S. Kumar$^{2,3}$,
G. Hosseinzadeh$^{4,5}$,
D. Hiramatsu$^{4,5}$,
I. Arcavi$^{4,5,6}$, \newauthor
D.~A. Howell$^{4,5}$,
S. Valenti$^{7}$,
D. J. Sand$^{8}$,
J.~T. Parrent$^{9}$,
C. McCully$^{4,5}$, \newauthor 
and A.~V. Filippenko$^{2,10}$ \\
$^{1}$ Department of Astronomy, University of Washington, Box 351580, Seattle, WA 98195-1580, USA \\
$^{2}$ Department of Astronomy, University of California, Berkeley, CA 94720-3411, USA \\
$^{3}$ Google Lick Predoctoral Fellow \\
$^{4}$ Las Cumbres Observatory, Goleta, CA 93117, USA \\
$^{5}$ Physics Department, University of California, Santa Barbara, CA 93106, USA \\
$^{6}$ Einstein Fellow \\
$^{7}$ Department of Physics,  University of California, Davis, 1 Shields Ave, Davis, CA 95616, USA \\
$^{8}$ Department of Astronomy and Steward Observatory, University of Arizona, 933 N Cherry Ave, Tucson, AZ 85719, USA \\
$^{9}$ Harvard-Smithsonian Center for Astrophysics, 60 Garden Street, Cambridge, MA 02138, USA \\
$^{10}$ Senior Miller Fellow, Miller Institute for Basic Research in Science, 
University of California, Berkeley, CA 94720, USA } 

\begin{document}
\pagerange{\pageref{firstpage}--\pageref{lastpage}} \pubyear{2016}

\maketitle

\label{firstpage}

\begin{abstract}

We present late-time spectra of eight Type Ia supernovae (SNe\,Ia) obtained at $>200$\,days after peak brightness using the Gemini South and Keck telescopes. All of the SNe\,Ia in our sample were nearby, well separated from their host galaxy's light, and have early-time photometry and spectroscopy from the Las Cumbres Observatory (LCO). Parameters are derived from the light curves and spectra such as peak brightness, decline rate, photospheric velocity, and the widths and velocities of the forbidden nebular emission lines. We discuss the physical interpretations of these parameters for the individual SNe\,Ia and the sample in general, including comparisons to well-observed SNe\,Ia from the literature. There are possible correlations between early-time and late-time spectral features that may indicate an asymmetric explosion, so we discuss our sample of SNe within the context of models for an offset ignition and/or white dwarf collisions. A subset of our late-time spectra are uncontaminated by host emission, and we statistically evaluate our nondetections of H$\alpha$ emission to limit the amount of hydrogen in these systems. Finally, we consider the late-time evolution of the iron emission lines, finding that not all of our SNe follow the established trend of a redward migration at $>200$\,days after maximum brightness. 

\end{abstract}


\section{Introduction} \label{sec:intro}

Type Ia supernovae (SNe\,Ia) are a powerful standardisable candle, but their progenitor scenarios and explosion mechanisms are not yet well understood (e.g., \citealt{2011NatCo...2E.350H,2014ARA&A..52..107M}). Nebular-phase spectra, taken $>200$\,days after peak brightness when the material is optically thin, are a unique diagnostic for SN\,Ia models --- especially when combined with time-series spectra and photometry starting before peak brightness. Modern wide-field surveys have increased the number of nearby SNe\,Ia that are sufficiently bright  for nebular-phase spectroscopy, and the new robotic spectrographs and imagers of the Las Cumbres Observatory (LCO; \citealt{2013PASP..125.1031B}) have improved our ability to monitor SNe from early times. In this paper we present and analyse a new sample of nebular-phase spectra of eight nearby SNe\,Ia obtained with programs at Keck and Gemini Observatories.

There are several ways that nebular spectra of SNe\,Ia can help to constrain the progenitor scenario and/or explosion mechanism. Once the material is optically thin, the spectrum is dominated by the broad, forbidden emission lines of the nucleosynthetic products of the thermonuclear explosion: nickel, cobalt, and iron. \cite{2007Sci...315..825M} use a line-emission code to model a set of SN\,Ia nebular spectra, measuring the masses of radioactive and stable nucleosynthetic products and intermediate-mass elements. They show that the mass ratios of these components vary continuously between SNe\,Ia but that all events have a similar total mass. Building on the work of \cite{1994ApJ...426L..89K}, \cite{2015MNRAS.454.3816C} established that the mass of $^{56}$Ni synthesised in the explosion can also be measured from the decline rate of [\ion{Co}{III}] $\lambda5893$ line flux in nebular-phase spectra. In the past, the 5893\,\AA\ feature has been suspected to have contributions from \ion{Na}{I}~D (e.g., \citealt{1994ApJ...426L..89K,2011MNRAS.410.1725T}), but \cite{2014MNRAS.439.3114D} have suggested that the sodium emission is weak compared to cobalt --- although at extremely late times ($\sim1000$\,days), after the $^{56}$Co has decayed, the \ion{Na}{I}~D line may be visible (e.g., \citealt{2015MNRAS.454.1948G}).

Aside from the total mass of nucleosynthetic products, the nebular-phase spectrum can also help reveal its distribution within the supernova (SN) ejecta (e.g., Figure 1 of \citealt{2007Sci...315..825M}). \cite{2010Natur.466...82M} suggested that an off-centre ignition followed by an asymmetric explosion causes the nebular emission lines to be redshifted or blueshifted, depending on the observer's viewing angle, and such asymmetries may be a contributing factor to the SN\,Ia width-luminosity correlation \citep{2011MNRAS.413.3075M}. Asymmetry is also a natural outcome of the collisional scenario in which two white dwarfs collide \citep{2009ApJ...705L.128R,2011A&A...528A.117P}, creating synthesised material with a bimodal velocity distribution that is observable as double-peaked nebular lines, such as those identified by \cite{2015MNRAS.454L..61D}. The leading progenitor scenarios for SNe\,Ia include accretion from or merger with another white dwarf (double-degenerate) or accretion from a main-sequence or red-giant star (single-degenerate). In the latter scenario, leftover hydrogen from the companion star might become visible as a narrow feature at late times after the flux of the SN subsides (e.g., \citealt{2005A&A...443..649M,2007ApJ...670.1275L}), but this has so far only been tentatively identified for one SN\,Ia by \cite{2016MNRAS.457.3254M}. 

Since SNe\,Ia fade by $\gtrsim5$\,mag between their light-curve peak and 200\,days later, nebular-phase spectroscopy is possible only for objects that are relatively nearby and thus reach brighter apparent peak magnitudes. Nearby SNe are also easier to monitor with a dense sampling of photometry and spectroscopy during the first few months. This is advantageous because there are several photospheric-phase characteristics of SNe\,Ia that correlate with nebular-phase observations in a physically meaningful way. One example is the synthesised mass of radioactive $^{56}$Ni, $M_{\rm ^{56}Ni}$, which is directly proportional to the peak luminosity \citep{1982ApJ...253..785A}.
$M_{\rm ^{56}Ni}$ can also be measured from nebular-phase spectra, as mentioned above, and is important because it is the physical origin of the width-luminosity relation by which SNe\,Ia are calibrated to events of standard brightness \citep{1993ApJ...413L.105P}. Another example is the rate of decline of the velocity of the \ion{Si}{II} $\lambda 6355$ line, $\dot{v}_{\rm Si~II}$, during the weeks after peak brightness. In the off-centre explosion model of \cite{2010Natur.466...82M}, measuring a high or low $\dot{v}_{\rm Si~II}$ depends on the observer's viewing angle and is thus also correlated with the nebular emission lines appearing to be redshifted or blueshifted, as mentioned above.

In this work, we present nebular-phase spectra for eight nearby SNe\,Ia, obtained during several semesters of a dedicated follow-up program at Gemini Observatory South and a transient follow-up program at Keck Observatory. We also present photospheric-phase characteristics derived from the photometric and spectroscopic monitoring program at the Las Cumbres Observatory. In Section \ref{sec:obs} we present our sample of SNe\,Ia and our observations. We analyse our nebular-phase spectra in Section \ref{sec:ana}, first describing how we measure properties of the emission lines and then providing a comprehensive overview of the particular attributes of each individual SN\,Ia. In Section \ref{sec:disc} we interpret our observations in terms of the progenitor scenarios and explosion models for SNe\,Ia discussed above, and we conclude in Section \ref{sec:conc}.

\section{Observations}\label{sec:obs}

Our sample of SNe\,Ia was chosen with the following criteria: (1) nearby ($<100$\,Mpc) and bright enough to observe at late times (i.e., $B\lesssim22$\,mag at $\geq200$\,days after peak); (2) significantly offset from the host galaxy, in regions with a low background surface brightness; (3) well observed at early times by the LCO (an optical light curve that covers peak brightness, and at least two epochs of optical photospheric-phase spectra). The last criterion ensures that we have physical properties of the explosion such as the nickel mass and photospheric velocity in order to properly interpret and contextualise our nebular spectra. 

In Table \ref{tab:objects}, we list the names and coordinates of these SNe\,Ia, their redshift and distance based on host-galaxy observations, and the Galactic line-of-sight extinction and reddening terms that we will apply to our observations. The photospheric-phase data for our targets are described in Section \ref{ssec:early}. We present our nebular-phase optical spectroscopy and imaging from Gemini Observatory in \ref{ssec:obs_GMOS} and our late-time spectra from Keck Observatory in \ref{ssec:obs_Keck}. Section \ref{sec:ana} provides an analysis of these data, including detailed discussions of each of our SNe\,Ia as individual objects.

\begin{table*}
\begin{tabular}{lccccc}
\hline
\hline
Name & SN Coordinates & Redshift  & Distance & \multicolumn{2}{c}{Galactic Extinction$\rm ^{(a)}$}  \\ 
           & RA, Dec (J2000)      & ($z$)       & (Mpc)     & $A_B$ (mag)  & $E(B-V)$               \\ 
\hline
SN\,2012fr       & 03:33:35.99 $-$36:07:37.7   & $0.005457 \pm 0.000003$$\rm ^{(b1)}$ & $18.14\pm2.76$$\rm ^{(d)}$ & 0.074  & $0.0177$  \\ 
SN\,2012hr      & 06:21:38.46 $-$59:42:50.6   & $0.007562 \pm 0.000013$$\rm ^{(b2)}$ & $40.0\pm6.2$$\rm ^{(d)}$     & 0.163  & $0.0389$  \\ 
SN\,2013aa      & 14:32:33.88 $-$44:13:27.8   & $0.003999 \pm 0.000007$$\rm ^{(b2)}$ & $16.9\pm2.9$$\rm ^{(e)}$     & 0.616  & $0.1458$  \\
SN\,2013cs      & 13:15:14.81 $-$17:57:55.6   & $0.009243 \pm 0.000003$$\rm ^{(b3)}$ & $37.9\pm2.7$$\rm ^{(c)}$    & 0.337  & $0.0788$  \\
ASASSN-14jg & 23:33:13.90 -60:34:11.5   & $0.014827\pm 0.000150$$\rm ^{(b4)}$  & $62.2\pm4.4$$\rm ^{(c)}$   & 0.055  & $0.0128$  \\
SN\,2015F        & 07:36:15.76 $-$69:30:23.0   & $0.004890\pm 0.000018$$\rm ^{(b7)}$  & $17.1\pm5.3$$\rm ^{(e)}$     & 0.735  & $0.1794$  \\
SN\,2013dy       & 22:18:17.60 +40:34:09.6 &  $0.003889 \pm 0.000017$$\rm ^{(b5)}$ & $12.35\pm2.34$$\rm ^{(d)}$ & 0.557  & $0.1320$  \\
SN\,2013gy       & 03:42:16.88 $-$04:43:18.5  & $0.014023 \pm 0.000020$$\rm ^{(b6)}$ & $59.4\pm4.2$$\rm ^{(c)}$   & 0.208  & $0.0485$  \\
\hline
\end{tabular}
\caption{For each of the SNe\,Ia in our sample, we list its name, coordinates, redshift and distance based on host-galaxy observations, and the Galactic line-of-sight extinction and reddening terms.  (a) Extinction values are from \protect\cite{2011ApJ...737..103S}. (b1) \protect\cite{1996ApJ...463...60B}. (b2) \protect\cite{2004AJ....128...16K}. (b3) \protect\cite{2011ApJS..197...28P}. (b4) \protect\cite{2009MNRAS.399..683J}. (b5) \protect\cite{1992ApJS...81....5S}. (b6) \protect\cite{2005AJ....130.1037C}. (b7) \protect\cite{2006MNRAS.371.1855W}. (c) The Galactocentric Hubble flow distance assuming $H_0=70$\,km\,s$^{-1}$\,Mpc$^{-1}$, $\Omega_M=0.3$, $\Omega_{\Lambda}=0.7$. (d) Combination of individually referenced distances to host galaxy on NED. (e) \protect\cite{Tully1988}. }
\label{tab:objects}
\end{table*}

\subsection{Early-Time Observations}\label{ssec:early}

\begin{figure*}
\begin{center}
\includegraphics[width=8cm]{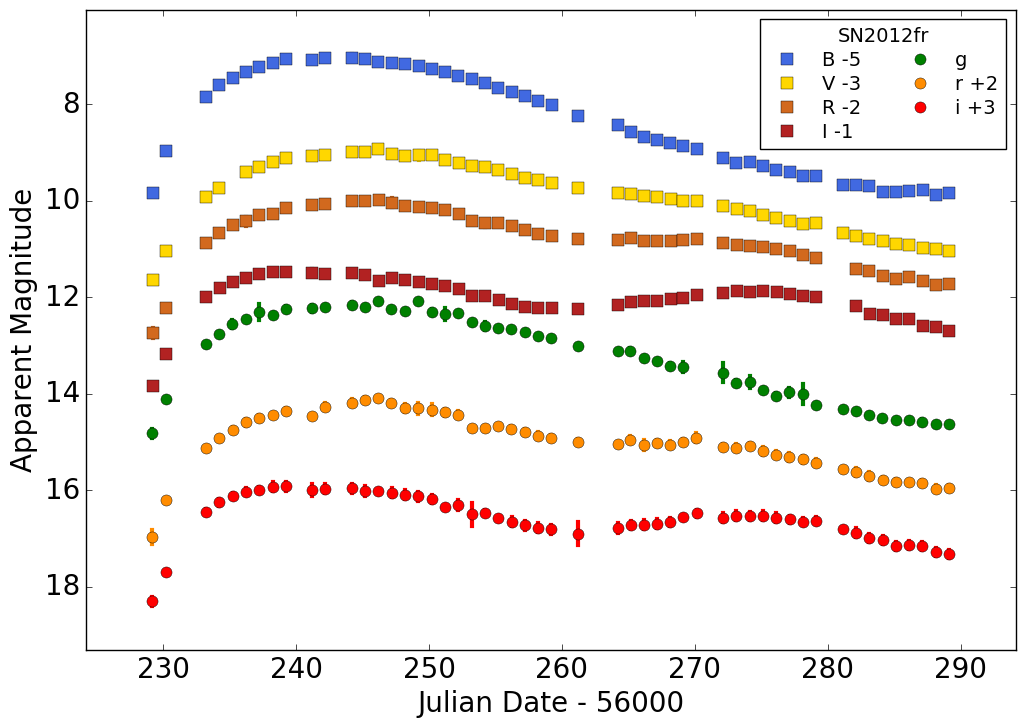}
\includegraphics[width=8cm]{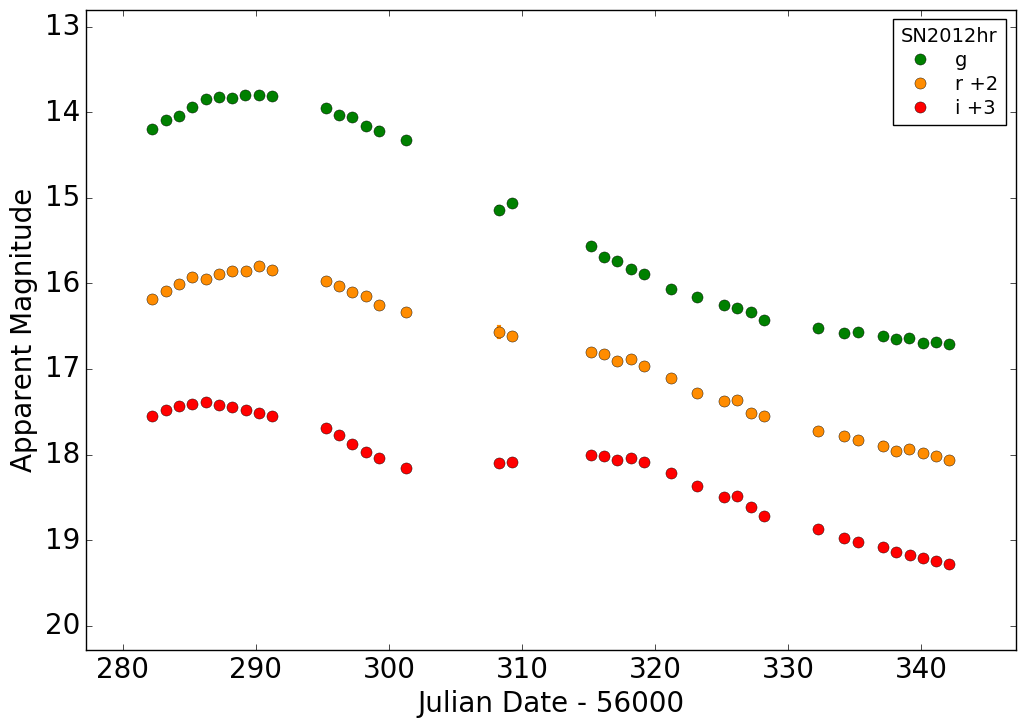}
\includegraphics[width=8cm]{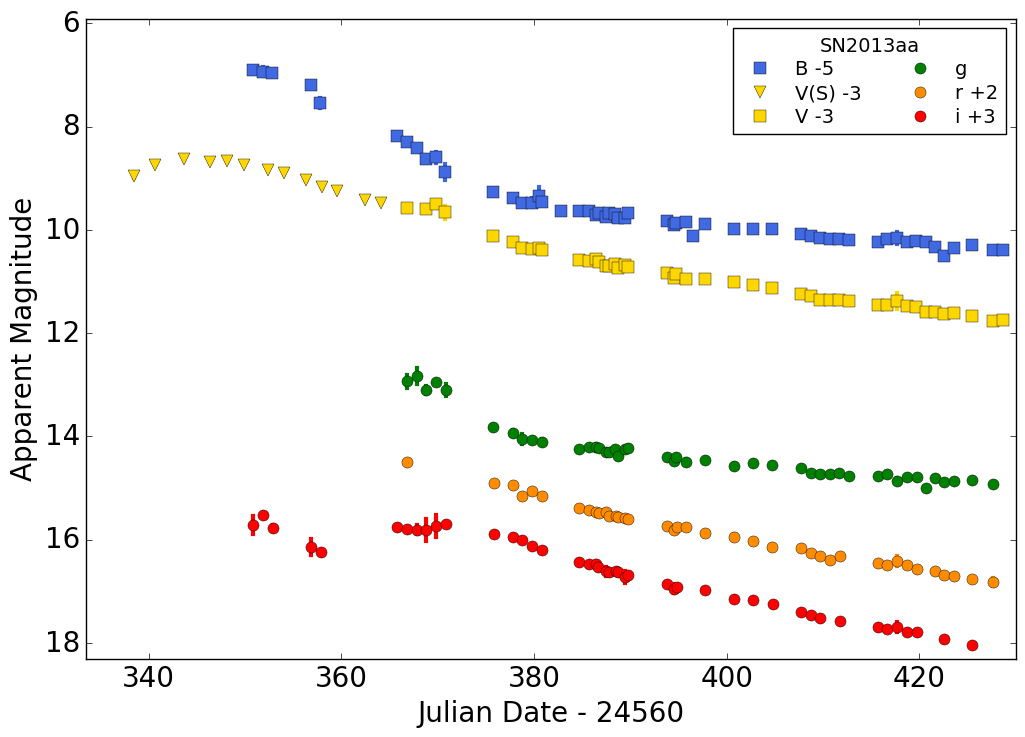}
\includegraphics[width=8cm]{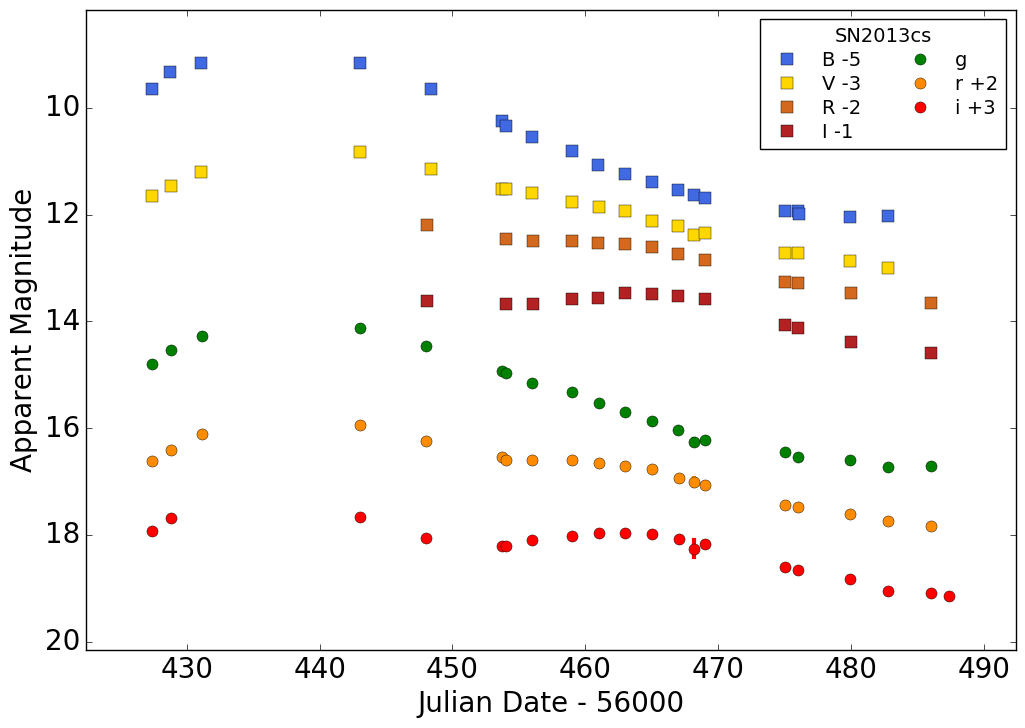}
\includegraphics[width=8cm]{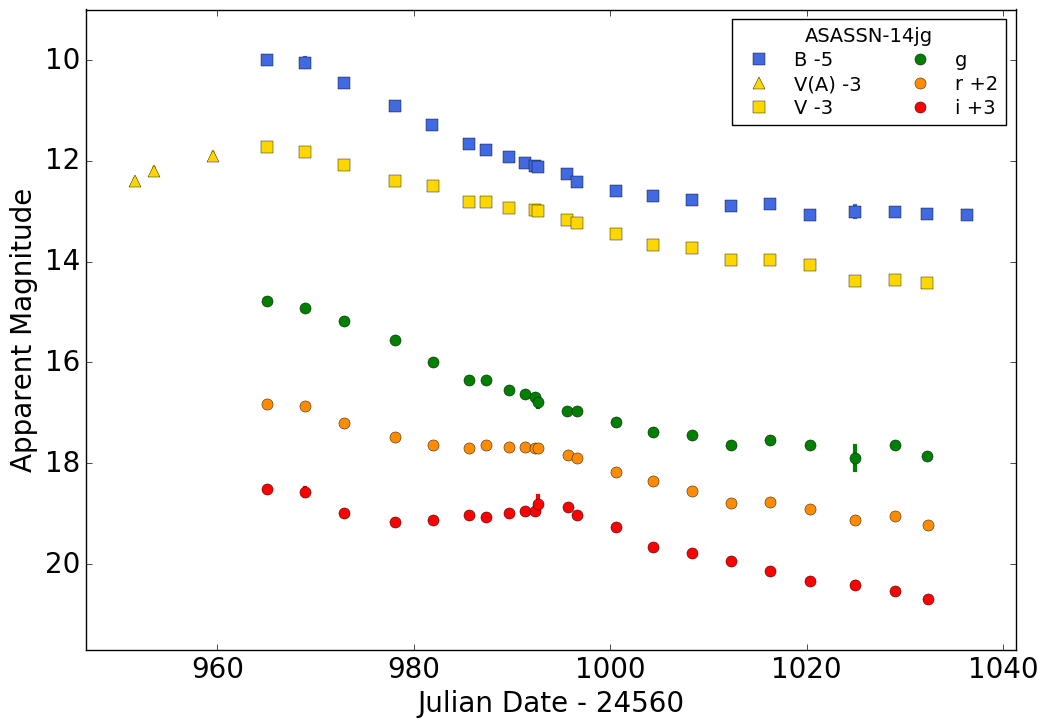}
\includegraphics[width=8cm]{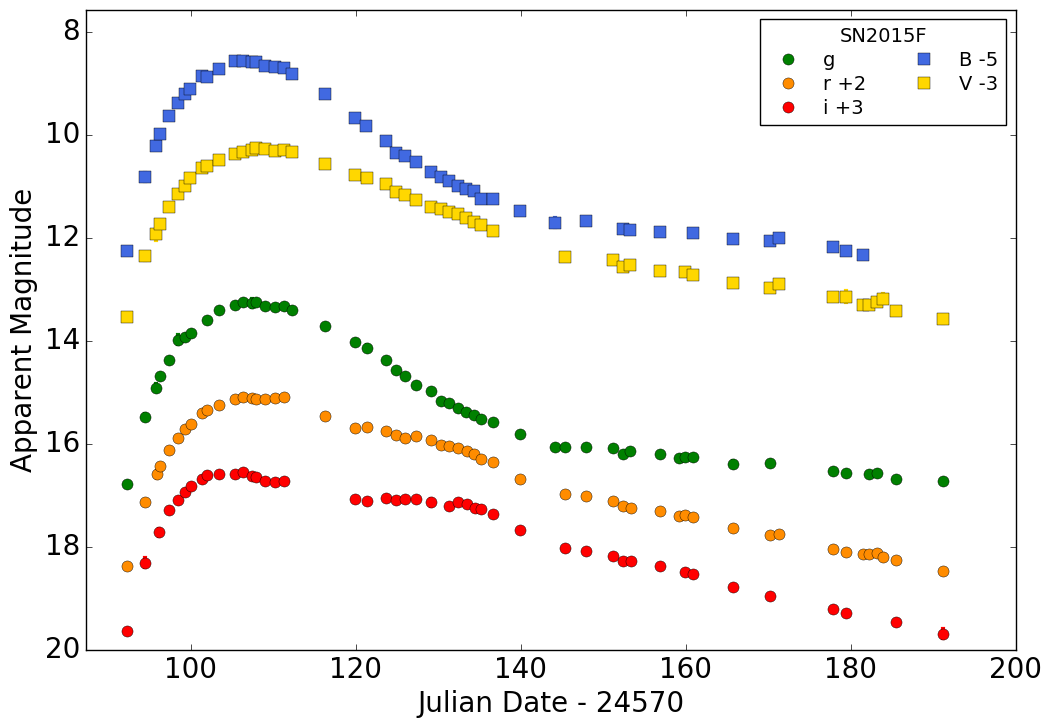}
\includegraphics[width=8cm]{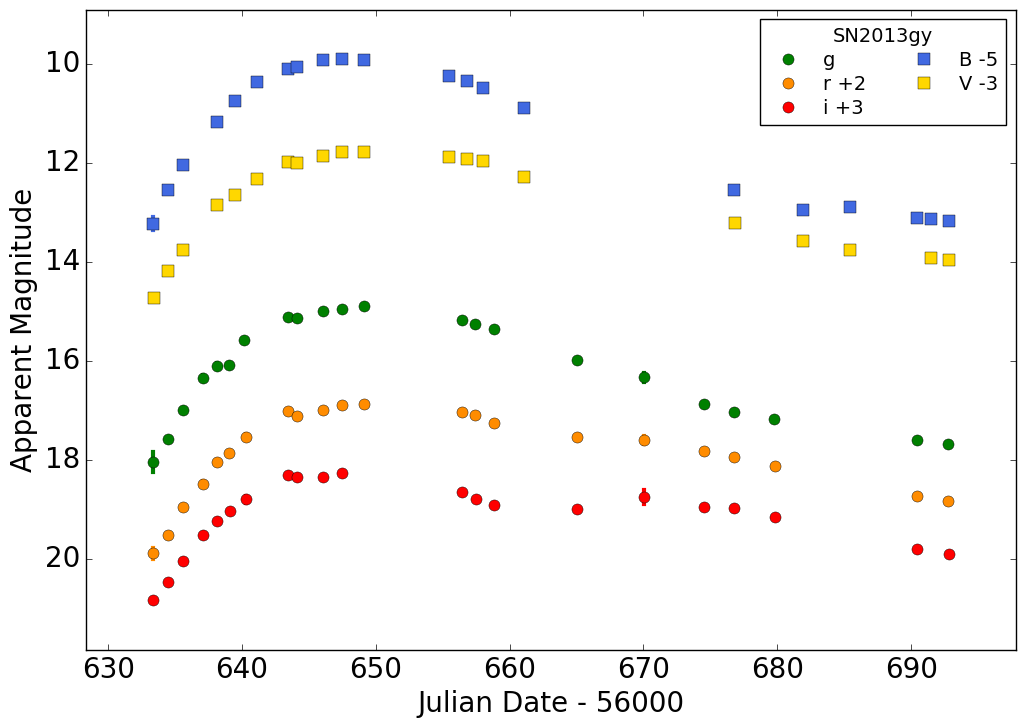}
\includegraphics[width=8cm]{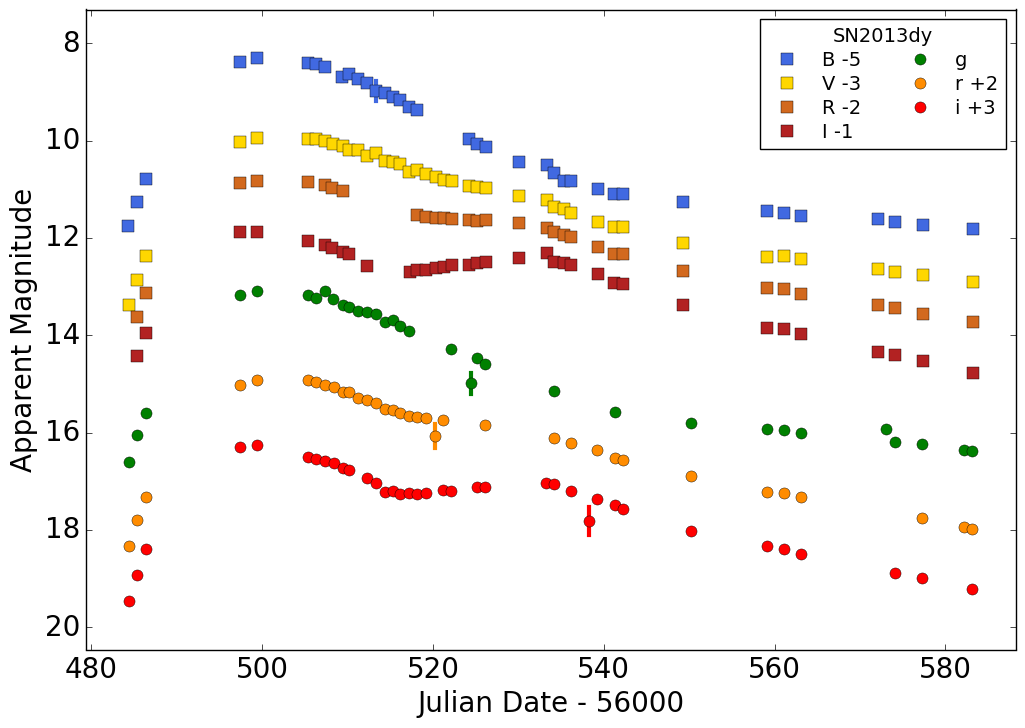}
\caption{Multiband SN light curves from Las Cumbres Observatory. Additional ASASSN and {\it Swift} photometry for ASASSN-14jg and SN\,2013aa, respectively, is included to extend the $V$-band light curves to premaximum epochs (yellow triangles labeled as $V$(A) and $V$(S); \protect\citealt{2014ATel.6637....1H}; \protect\citealt{2014Ap&SS.354...89B}). \label{fig:lc}}
\end{center}
\end{figure*}

At early times, our targeted SNe were monitored with optical imaging and spectroscopy as part of the Supernova Key Project by the Las Cumbres Observatory (LCO). Typical follow-up observations for nearby SNe include imaging in the $BVgri$ filters every 3--7\,days and low-resolution spectroscopy with the FLOYDS instrument every 5--10\,days. These data are all automatically processed and calibrated using in-house pipelines designed by LCO astronomers (see Appendix B of \citealt{2016MNRAS.459.3939V}). We display the multiband light curves from LCO in Figure \ref{fig:lc} and provide the photometry in Table \ref{tab:lc}. The photometry for SN\,2015F was previously published by \cite{2017MNRAS.464.4476C}. We use the {\sc Python} package {\sc SNCosmo}\footnote{http://sncosmo.readthedocs.org/} to fit light curves to the photometry, and the resulting peak apparent magnitudes and post-peak decline rates $\Delta m_{15} (B)$ are listed in Table \ref{tab:lcfit}. Unfortunately, most of LCO's near-peak photometry of SN\,2013aa was saturated, so for its light-curve fit we used $V$-band photometry from the $Swift$ Optical/Ultraviolet Supernova Archive (SOUSA\footnote{$V$-band photometry was derived from observations with the Ultraviolet/Optical Telescope (UVOT; \citealp{2005SSRv..120...95R}) on the {\it Swift} spacecraft \citep{2004ApJ...611.1005G}. Reduced and calibrated photometry was downloaded from the SOUSA website at \url{http://swift.gsfc.nasa.gov/docs/swift/sne/swift\_sn.html}. The reduction is based on that of \citet{2009AJ....137.4517B}, including subtraction of the host-galaxy count rates, and uses the revised UV zeropoints and time-dependent sensitivity from \citet{2011AIPC.1358..373B}.}; \citealt{2014Ap&SS.354...89B}), which is shown also for reference in Figure \ref{fig:lc}.

Given the established correlation between the \ion{Si}{II} velocity gradient in the weeks after peak brightness ($\dot{v}_{\rm Si~II}$) and the nebular-phase emission-line velocity (e.g., \citealt{2010Natur.466...82M}), we attempt to classify our SNe\,Ia as either ``low" or ``high" velocity gradient (LVG or HVG). For our SNe\,Ia with at least two spectra between days $0$ and $14$, we make a simple linear fit to the velocity of the \ion{Si}{II} $\lambda 6355$ line from the LCO FLOYDS spectra (i.e., SNe\,2012fr, 2012hr, 2013aa, and 2013dy in Table \ref{tab:vSiII}). The true decline is not linear, but this is a fine approximation for our purposes of classifying as LVG or HVG. Of these, only SN\,2012hr approaches a value of $\dot{v}_{\rm Si~II}$ that is HVG-like. Unfortunately, not all of our SNe\,Ia had the necessary photospheric-phase spectral coverage for this, but we can still make an educated guess at whether they are LVG or HVG-like SNe\,Ia. \cite{2005ApJ...623.1011B} established that LVG SNe\,Ia typically have $\dot{v}_{\rm Si~II} < 70$\,km\,s$^{-1}$\,d$^{-1}$ and $v_{\rm Si~II}({\rm peak}) \lesssim 12,000$\,km\,s$^{-1}$, and that HVG SNe\,Ia typically have $\dot{v}_{\rm Si~II} > 70$\,km\,s$^{-1}$\,d$^{-1}$ and $v_{\rm Si~II}({\rm peak}) \gtrsim 12,000$\,km\,s$^{-1}$. Furthermore, \cite{2013MNRAS.430.1030S} show that ${v}_{\rm Si~II}$ can be used as a proxy for $\dot{v}_{\rm Si~II}$ (in their Figure 6). For the remaining SNe\,Ia in our sample, we measure $v_{\rm Si~II}$ in the spectrum nearest to peak light (0, +12, $-7$, and $-1$\,days for SNe\,2013cs, 2014jg, 2015F, and 2013gy, respectively; Table \ref{tab:vSiII}), and use these ranges from \cite{2005ApJ...623.1011B} to estimate the velocity-gradient classification from a single epoch alone. Of these, only SN\,2013cs exhibits a velocity in the range of HVG-like SNe\,Ia. 

In Table \ref{tab:vSiII} we list the LVG and HVG determination for each of our SNe\,Ia, with a ``comments'' column describing the reasoning behind each determination. To summarize, we (or previous publications) have directly measured $\dot{v}_{\rm Si~II}$ for SNe\,Ia 2012fr, 2012hr, 2013aa, and 2013dy. Of these four, only SN\,2012hr has $\dot{v}_{\rm Si~II} > 70$\,km\,s$^{-1}$\,d$^{-1}$, but since there is a large uncertainty in our measurement of $\dot{v}_{\rm Si~II}$, and because SN\,2012hr also exhibits $v_{\rm Si~II}({\rm peak}) \lesssim 12,000$\,km\,s$^{-1}$ (like a LGV SN), its status as an HVG SN is not secure and we therefore list it as ``HVG?". We consider the other three SNe\,Ia (SNe\,2012fr, 2013aa, and 2013dy) securely classified as LVG, and list them as such. For the remaining four SNe\,Ia in our sample (SNe\,2013cs, ASASSN-14jg, 2015F, and 2013gy), we must instead rely on an indirect estimate of its LVG/HVG classification using $v_{\rm Si~II}({\rm peak})$ as a proxy for $\dot{v}_{\rm Si~II}$, as described in the previous paragraph. Of these, only SN\,2013cs has $v_{\rm Si~II}({\rm peak}) \gtrsim 12,000$\,km\,s$^{-1}$. However, we know from SN\,2012fr -- which exhibited $v_{\rm Si~II}({\rm peak}) \gtrsim 12,000$\,km\,s$^{-1}$ but was clearly an LVG SN\,Ia -- that the value of $v_{\rm Si~II}({\rm peak})$ is not a robust proxy for the velocity gradient. We therefore list SN\,2013cs as ``HVG?", and furthermore list the remaining three SNe\,Ia (ASASSN-14jg, SN\,2015F, and SN\,2013gy) as ``LVG?" to represent the uncertainty in using $v_{\rm Si~II}({\rm peak})$ instead of $\dot{v}_{\rm Si~II}$. Since our LVG and HVG assignments are more tentative estimates than secure measurements, in this work we will refrain from drawing any strong conclusions that rely on these classifications.

\begin{table}
\begin{centering}
\begin{tabular}{lccc}
\hline
\hline
SN Name & Filter & Modified JD & Magnitude \\
\hline
SN\,2012fr & $g$ & 56229.179 & $14.82\pm0.10$   \\
SN\,2012fr & $r$  & 56229.185 & $14.96\pm0.14$  \\
SN\,2012fr & $i$  & 56229.190 & $15.30\pm0.09$   \\
SN\,2012fr & $B$ & 56229.201 & $14.84\pm0.01$  \\
SN\,2012fr & $V$ & 56229.207 & $14.64\pm0.04$  \\
SN\,2012fr & $R$ & 56229.212 & $14.74\pm0.10$   \\
SN\,2012fr & $I$  & 56229.218  & $14.83\pm0.02$  \\
\hline
\end{tabular}
\caption{A sample of the photometry for our SNe\,Ia from LCO, in AB magnitudes. The full dataset is available online.}
\label{tab:lc}
\end{centering}
\end{table}

\begin{table*}
\begin{tabular}{l cc cc cc}
\hline
\hline
SN Name     & $t_{\rm peak}$  & $m_{\rm peak}(B)$  & $\Delta m_{15}(B)$ & $M_{\rm peak}(B)$  \\
 & [MJD] & mag & mag & mag \\
\hline
SN\,2012fr    & $56244.7 \pm 0.2$ & $11.93 \pm 0.03$ & $0.93 \pm 0.02$ & $-19.4 \pm 0.4$  \\
SN\,2012hr    & $56289.2 \pm 0.1$ & $13.75 \pm 0.03$ & $1.04 \pm 0.01$ & $-19.4 \pm 0.4$  \\
SN\,2013aa    & $56343.7 \pm 0.3$ & $11.61 \pm 0.03$ & $1.02 \pm 0.05$ & $-20.1 \pm 0.4$  \\
SN\,2013cs    & $56437.5 \pm 0.9$ & $14.23 \pm 0.03$ & $0.81 \pm 0.18$ & $-19.0 \pm 0.2$  \\
ASASSN-14jg   & $56961.6 \pm 0.2$ & $14.68 \pm 0.03$ & $0.92 \pm 0.01$ & $-19.3 \pm 0.2$  \\
SN\,2015F     & $57106.9 \pm 0.1$ & $13.48 \pm 0.03$ & $1.18 \pm 0.02$ & $-18.4 \pm 0.8$  \\
SN\,2013dy$^{(a)}$  & $56501.1$ & $13.23\pm0.01$ & $0.92\pm0.01$ & $-19.66\pm0.02$ \\
SN\,2013gy    & $56648.5 \pm 0.1$ & $14.95 \pm 0.03$ & $1.20 \pm 0.01$ & $-19.1 \pm 0.2$  \\
\hline
\end{tabular}
\caption{Light-curve fit parameters for our sample of SNe\,Ia. Peak absolute $B$-band magnitudes are calculated using the apparent $B$-band magnitudes and the distance modulii listed in Table \ref{tab:objects}, except for SN\,2013dy. (a) The values for SN\,2013dy come from \protect\cite{2015MNRAS.452.4307P}, who found that the distance derived from the light-curve fit is significantly farther than the reported Tully-Fisher distance for the host galaxy. The value of $M_{\rm peak}(B)$ in this table is derived from $m_{\rm peak}(B)=13.23$\,mag, $\mu_{\rm SN\,2013dy} = 31.49$\,mag, and $A_{B,{\rm host}} = 0.846$\,mag, which are in Section 3.1 of \protect\cite{2015MNRAS.452.4307P}, and also the line-of-sight extinction from the Milky Way as reported in Table \ref{tab:objects}. \protect\cite{2015MNRAS.452.4307P} do not provide an estimate of the uncertainty on the time of maximum brightness.}
\label{tab:lcfit}
\end{table*}

\begin{table*}
\begin{tabular}{llcccl}
\hline
\hline
SN Name        & Phases of LCO      &  $v_{\rm Si~II}({\rm peak})$     & $\dot{v}_{\rm Si~II}$            & Group & Comments \\
                        & FLOYDS Spectra [d]  &  [km\,s$^{-1}$]                  & [km\,s$^{-1}$\,d$^{-1}$]   &            &                    \\
\hline
SN\,2012fr       & -14, -11, -9, -8, -6, -5,        & 12,000     & $\sim 0$              & LVG      & See also \protect\cite{2013ApJ...770...29C} \\
& 0, 2, 4, 8, 10 & & & & \\
SN\,2012hr      & -7, 5, 15                             &  11,500     & $120 \pm 80$    & HVG?   & $\dot{v}_{\rm Si~II}({\rm peak})$ is HVG-like; $v_{\rm Si~II}({\rm peak})$ nearly HVG-like \\
SN\,2013aa     &  -3, -2, 3, 12, 13, 25, 26     &  10,200    & $50 \pm 10$      & LVG      & $\dot{v}_{\rm Si~II}({\rm peak})$ is LVG-like; $v_{\rm Si~II}({\rm peak})$ is LVG-like \\
SN\,2013cs      &  -1, 0                                  &  12,500    & \dots                    & HVG?  & $v_{\rm Si~II}({\rm peak})$ is HVG-like, but also 2012fr-like \\
ASASSN-14jg  & 12, 13                                &   10,200   & \ldots                   & LVG?      & $v_{\rm Si~II}({\rm peak})$ is LVG-like \\
SN\,2015F       &  -15, -11, -7                        &   10,100   & \ldots                   & LVG?      & $v_{\rm Si~II}({\rm peak})$ is LVG-like \\
SN\,2013dy     &  -10, -9, -7, +4, +6, +8,       &  10,300    & $25 \pm 20$      & LVG      & consistent with \protect\cite{2015MNRAS.452.4307P} \\
&  +10, +12, +20, +22, +32  & & & & \\
SN\,2013gy     & -14, -13, -6, -4, -1               &  10,700    & \ldots                   & LVG?      & $v_{\rm Si~II}({\rm peak})$ is LVG-like \\
\hline
\end{tabular}
\caption{Parameters of the photospheric-phase spectra (note that $\dot{v}_{\rm Si~II}$ is a positive number but represents the decline in velocity in km\,s$^{-1}$\,d$^{-1}$) and the derived \ion{Si}{II} velocity-gradient classifications from LCO FLOYDS observations --- except SN\,2013dy, for which we use the information previously published by \protect\cite{2015MNRAS.452.4307P}. }
\label{tab:vSiII}
\end{table*}

\subsection{Nebular-Phase Spectroscopy and Imaging from Gemini Observatory} \label{ssec:obs_GMOS}

Nebular spectra for SNe\,Ia in the southern hemisphere were obtained using the Gemini Multi-Object Spectrograph (GMOS; \citealt{1997SPIE.2871.1099D}) on the Gemini South telescope on Cerro Pachon, Chile. Our GMOS configurations use the 0.75\arcsec\ long slit, the B600 grating with no filter and a central wavelength of 4500\,\AA, and the R400 grating at a central wavelength of 7500\,\AA\ with the OG515 long-pass filter to block second-order light at $\lambda < 5200$\,\AA. These spectra are listed in Table \ref{tab:specobs}. We processed the data with the Gemini {\sc IRAF} package: the raw two-dimensional (2D) spectra were flat-fielded, bias-subtracted, and trimmed; cosmic-ray rejection was performed; wavelength calibration was determined using the Cu-Ar comparison arc lamp; and sky-subtracted one-dimensional (1D) apertures were extracted from the processed frames. The sensitivity function was determined from standard-star observations obtained in our GMOS configuration, as near in time as possible (often the same night), and applied to flux calibrate our data. We also used the standard-star spectra to remove telluric features from our spectra. 

The second-epoch spectra that we obtained of SN\,2012hr and SN\,2013aa turned out to have such low signal-to-noise ratios (S/N) that we excluded them from our analysis, but as they are at uniquely late times ($>450$\,days) we are publishing them to make them available to the community.

\begin{table*}
\begin{tabular}{lcccccc}
\hline
\hline
Object & Observation & Phase & Instrument & Grating and Central & Filter & Exposure  \\ 
            & Date (UT)    & [days] &                    & Wavelength [nm]     &          & Time [s]    \\
\hline
SN\,2012fr         & 08-29-2013 & +289   & GMOS & B600/450.0 & none      & 3$\times$400      \\ 
SN\,2012fr         & 08-29-2013 & +289   & GMOS & R400/750.0 & OG515  & 3$\times$600      \\ 
SN\,2012fr         & 01-02-2014 & +415   & GMOS & B600/450.0  & none     & 3$\times$1800     \\ 
SN\,2012fr         & 01-02-2014 &  +415  & GMOS & R400/750.0 & OG515  & 4$\times$1800    \\ 
SN\,2012hr$^{(a)}$  & 10-06-2013 & +283   & GMOS & B600/450.0  & none     & 4$\times$800      \\ 
SN\,2012hr$^{(a)}$  & 10-06-2013 &  +283  & GMOS & R400/750.0 & OG515   & 4$\times$1200    \\ 
SN\,2012hr$^{(b)}$  & 03-29-2014 &  +457  & GMOS & B600/450.0  & none      & 3$\times$2400  \\
SN\,2013aa        & 03-31-2014 & +400   & GMOS & B600/450.0 & none      & 3$\times$600      \\ 
SN\,2013aa        & 03-27-2014 &  +396  & GMOS & R400/750.0 & OG515  & 3$\times$900      \\ 
SN\,2013aa$^{(b)}$ & 07-02-2014 &  +493  & GMOS & B600/450.0 & none      & 3$\times$600     \\
SN\,2013aa$^{(b)}$  & 07-02-2014 &  +493  & GMOS & R400/750.0 & OG515  & 3$\times$900   \\
SN\,2013cs        & 02-18-2014 & +265   & GMOS & B600/450.0 & none      & 3$\times$600      \\ 
SN\,2013cs        & 02-11-2014 &  +258  & GMOS & R400/750.0 & OG515   & 3$\times$900      \\ 
ASASSN-14jg  & 07-25-2015 & +267   & GMOS & B600/450.0 & none      & 4$\times$1200     \\
ASASSN-14jg  & 07-27-2015 & +269  & GMOS & R400/750.0 & OG515   & 8$\times$1200     \\
SN\,2015F         & 12-30-2015 & +280   & GMOS & B600/450.0 & none      & 3$\times$800       \\ 
SN\,2015F         & 01-01-2016 & +282   & GMOS & R400/750.0 & OG515   & 4$\times$900      \\ 
\hline
\end{tabular} 
\caption{Nebular-phase spectroscopic observations from Gemini Observatory obtained by our program. (a) Published by \protect\cite{2015MNRAS.454.3816C}. (b) Spectra having poor S/N; not included in the analysis. }
\label{tab:specobs}
\end{table*}

We also included a sequence of $\sim 30$\,s $gri$ images with each GMOS observation, which typically occurred during target acquisition. We reduced these images using the Gemini {\sc IRAF} package, and calibrated the photometry using the sequence of local standards used for the LCO imaging analysis. All of our GMOS spectra were flux calibrated to match the late-time photometry, and the dates and magnitudes are listed in Table \ref{tab:GMOSphot}. In general, our observing requirements were relaxed to include poor weather conditions, particularly for semesters during which our program was in Gemini's Band 3; this is the cause for the variety in photometric uncertainties. 

\begin{table}
\begin{centering}
\begin{tabular}{lcl}
\hline
\hline
SN Name & Date & Photometry \\
\hline
SN\,2012fr   & 2013-08-29  & $g=18.2 \pm 0.3$ \\
                    &                      & $i=19.3\pm0.1$    \\
                    & 2014-01-03   & $g=20.2\pm0.2$   \\
                    &                      & $r=22.0\pm0.1$     \\
                    &                      & $i=20.65\pm0.04$  \\
SN\,2012hr  & 2013-10-06   & $g=20.18\pm0.03$ \\
                    &                      & $r=21.71\pm0.06$  \\
                    &                      & $i=20.78\pm0.05$  \\
                    & 2014-03-29   & $g=22.86\pm0.08$ \\
SN\,2013aa  & 2014-02-28  & $g=19.3\pm0.2$ \\
                    &                      & $r=21.0\pm0.1$ \\
                    &                      & $i=20.0\pm0.3$ \\
                    & 2014-03-27  & $g=19.95\pm0.05$ \\
                    &                      & $r=21.59\pm0.07$ \\
                    &                      & $i=20.08\pm0.07$ \\
SN\,2013cs  & 2014-02-11  & $g=20.02\pm0.06$  \\
                    &                      & $r=21.50\pm0.08$ \\
                    &                      & $i=20.77\pm0.05$ \\
ASASSN-14jg  & 2015-07-25 & $g=20.3\pm0.2$ \\
                    &                      & $r=20.8\pm0.3$ \\
                    &                      & $i=20.6\pm0.2$ \\
SN\,2015F   & 2015-12-30  & $g=20.06\pm0.09$ \\
                    &                      & $r=21.7\pm0.1$ \\ 
                    &                      & $i=20.37\pm0.05$ \\
\hline
\end{tabular}
\caption{Photometry from GMOS imaging. Photometry with large uncertainties can be attributable to the fact that our observing requirements were relaxed to include poor conditions.}
\label{tab:GMOSphot}
\end{centering}
\end{table}

\subsection{Nebular-Phase Spectroscopy from Keck Observatory} \label{ssec:obs_Keck}

Nebular spectra for SNe\,Ia in the northern hemisphere were obtained with the Low Resolution Imaging Spectrometer (LRIS; \citealt{1995PASP..107..375O}) at Keck Observatory. We used the 1.0\arcsec\ slit rotated to the parallactic angle to minimise the effects of atmospheric dispersion (\citealt{1982PASP...94..715F}; in addition, LRIS has an atmospheric dispersion corrector). In our LRIS configuration, coverage in the blue with the 600/4000 grism extends over $\lambda=3200$--5600\,\AA\ with a dispersion of 0.63\,\AA\,pixel$^{-1}$ and a full width at half-maximum intensity (FWHM) resolution of $\sim4$\,\AA. We use the 5600\,\AA\  dichroic, and our coverage in the red with the 400/8500 grating extends over $\lambda=5600$--10,200\,\AA\  with a dispersion of 1.16\,\AA\,pixel$^{-1}$ and a resolution of $\rm FWHM \approx 7$\,\AA. These spectra are listed in Table \ref{tab:keckspecobs}. Note that one Keck spectrum was obtained with DEIMOS; see Pan et al. (2015) for details.

The data were reduced with routines written specifically for LRIS in the Carnegie {\sc Python} ({\sc CarPy}) package. The 2D images were flat-fielded, corrected for distortion along the $y$ (slit) axis, wavelength calibrated with arc-lamp spectra, and cleaned of cosmic rays before extracting the 1D spectrum of the target. These spectra were flux calibrated using a sensitivity function derived from observations of a standard star obtained the same night in the same instrument configuration. The standard-star spectrum was also used to remove the telluric sky absorption features.

\begin{table*}
\begin{tabular}{lccccccl}
\hline
\hline
Object & Observation & Phase & Instrument & Grism or Grating & Slit      & Exposure & Published \\ 
           & Date (UT)     & [days] &                   &                              & Width & Time [s]   &                   \\
\hline 
SN\,2013dy & 2014-06-26 & $\sim335$ & DEIMOS & 600ZD                & 0.8\arcsec & $1 \times 1200$ & \cite{2015MNRAS.452.4307P} \\ 
SN\,2013dy & 2014-09-24 & $\sim425$ & LRIS & 600/4000, 400/8500 & 1.0\arcsec & $2 \times 1000$ & \cite{2015MNRAS.452.4307P} \\
SN\,2013dy & 2014-11-20 & $\sim482$ & LRIS & 600/4000, 400/8500 & 1.0\arcsec & $2 \times 1000$ & \cite{2015MNRAS.452.4307P} \\
SN\,2013gy & 2014-09-24 & $\sim277$ & LRIS & 600/4000, 400/8500 & 1.0\arcsec & $2 \times 1200$  & \cite{2015MNRAS.454.3816C} \\
SN\,2013gy & 2015-02-20 & $\sim426$ & LRIS & 600/4000, 400/8500 & 1.0\arcsec & $2 \times 1200$  & \\
\hline
\end{tabular} 
\caption{Nebular-phase spectroscopic observations from Keck Observatory, some of which have been previously published.}
\label{tab:keckspecobs}
\end{table*}

\section{Analysis}\label{sec:ana}

\begin{figure*}
\begin{center}
\includegraphics[width=8cm]{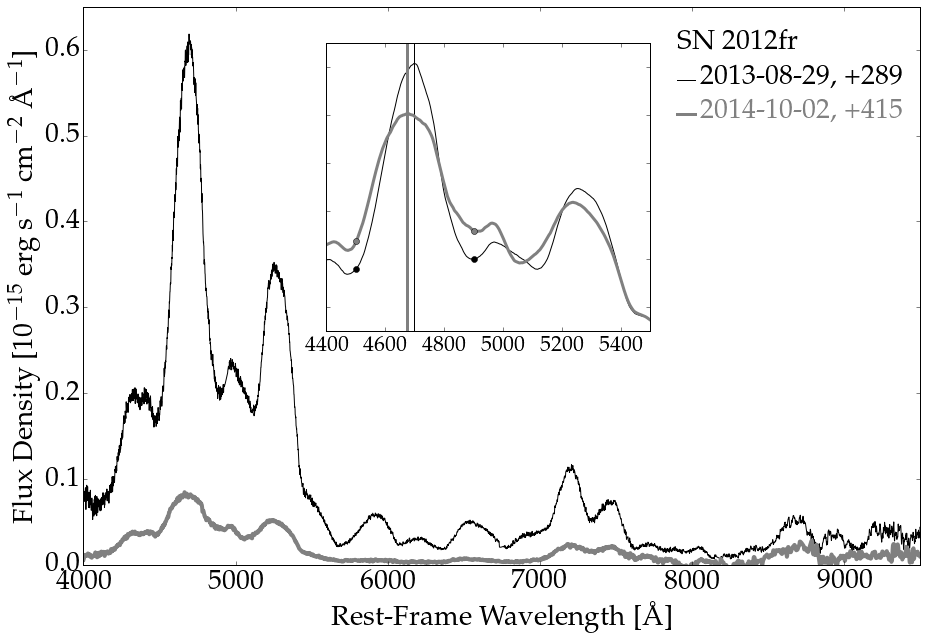}
\includegraphics[width=8cm]{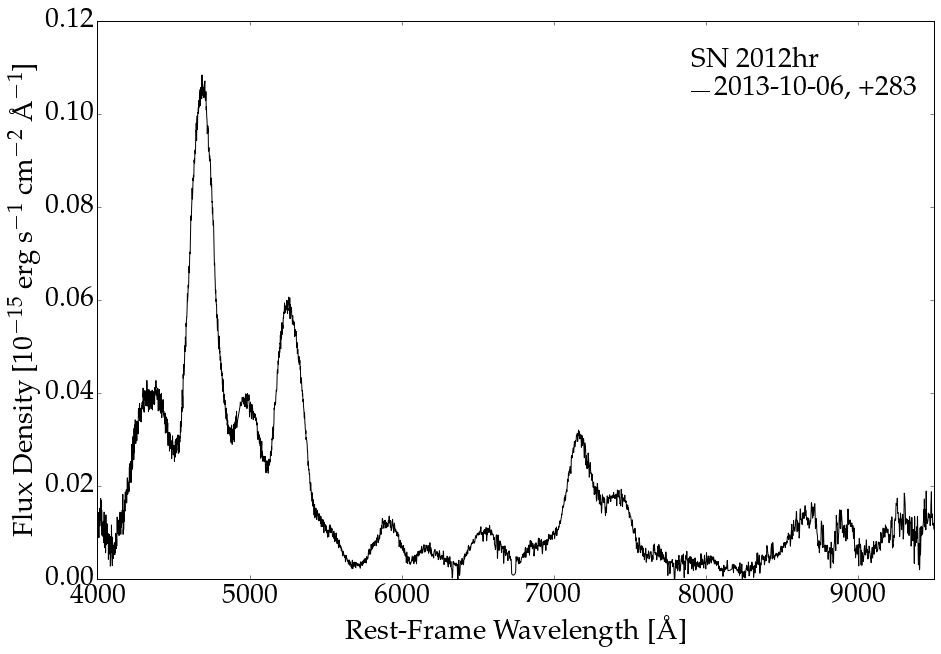}
\includegraphics[width=8cm]{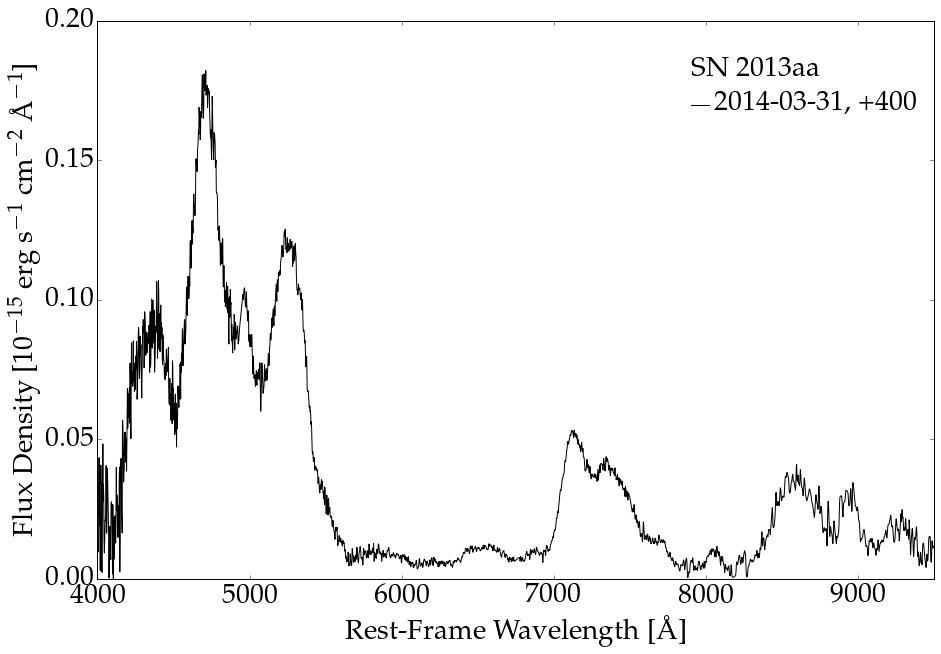}
\includegraphics[width=8cm]{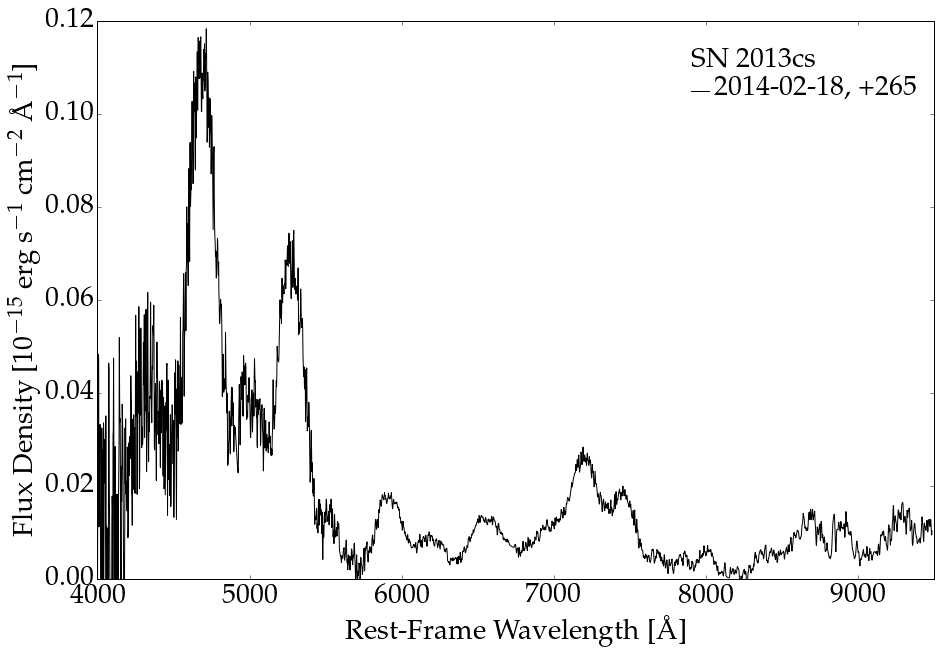}
\includegraphics[width=8cm]{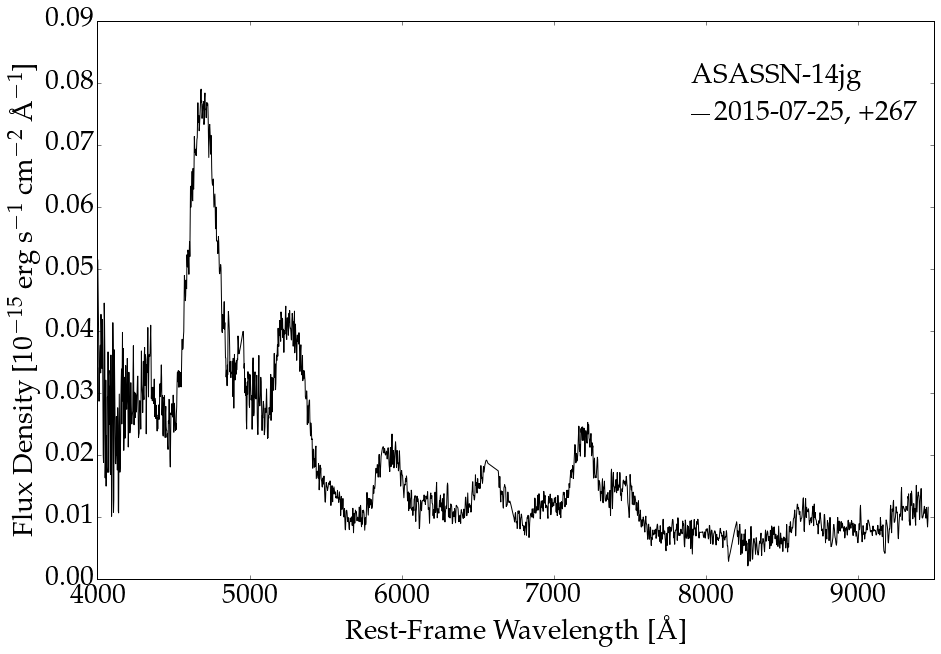}
\includegraphics[width=8cm]{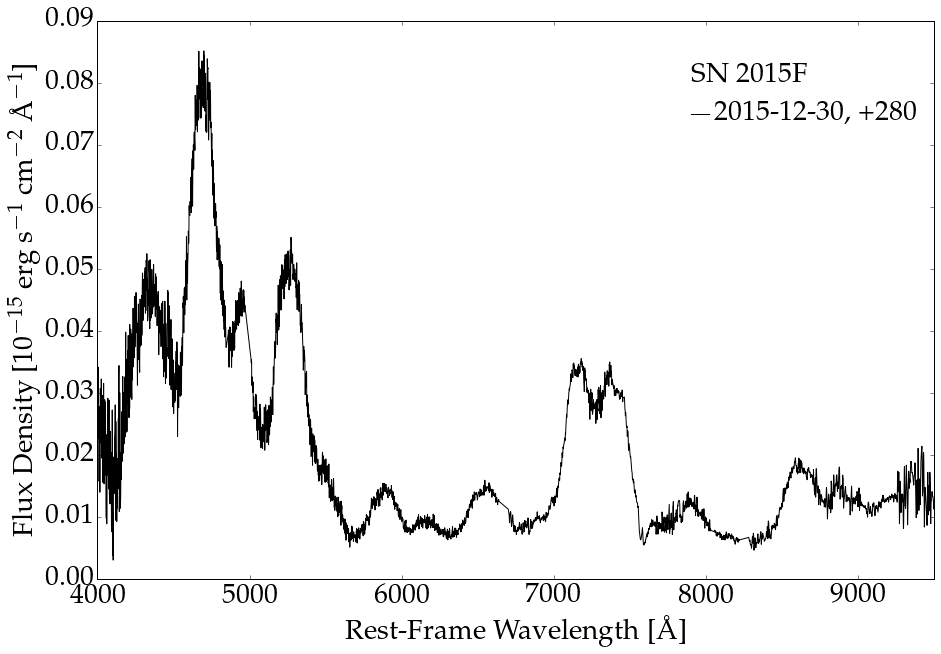}
\includegraphics[width=8cm]{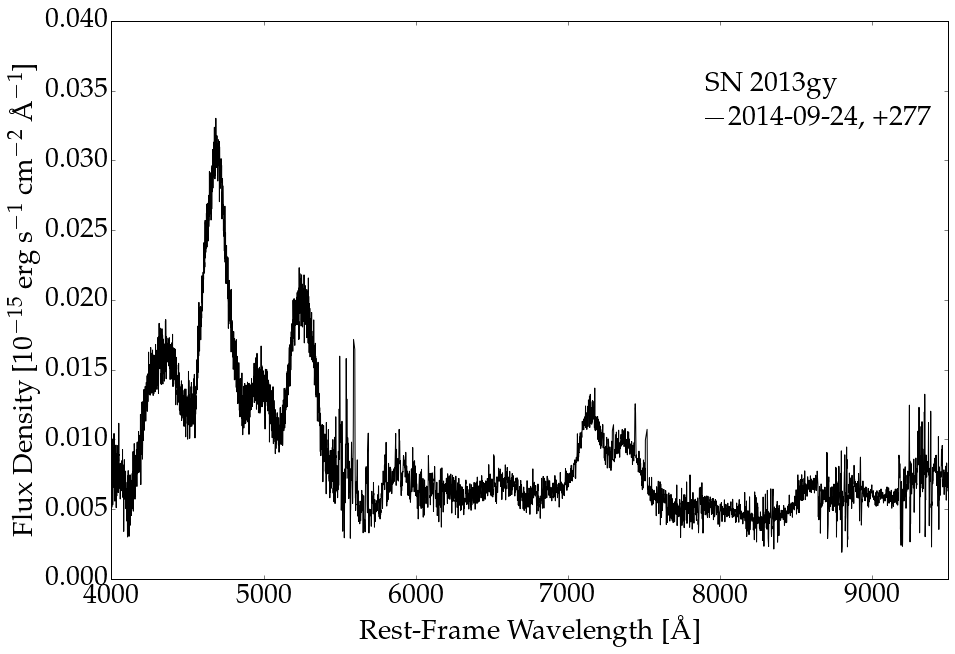}
\includegraphics[width=8cm]{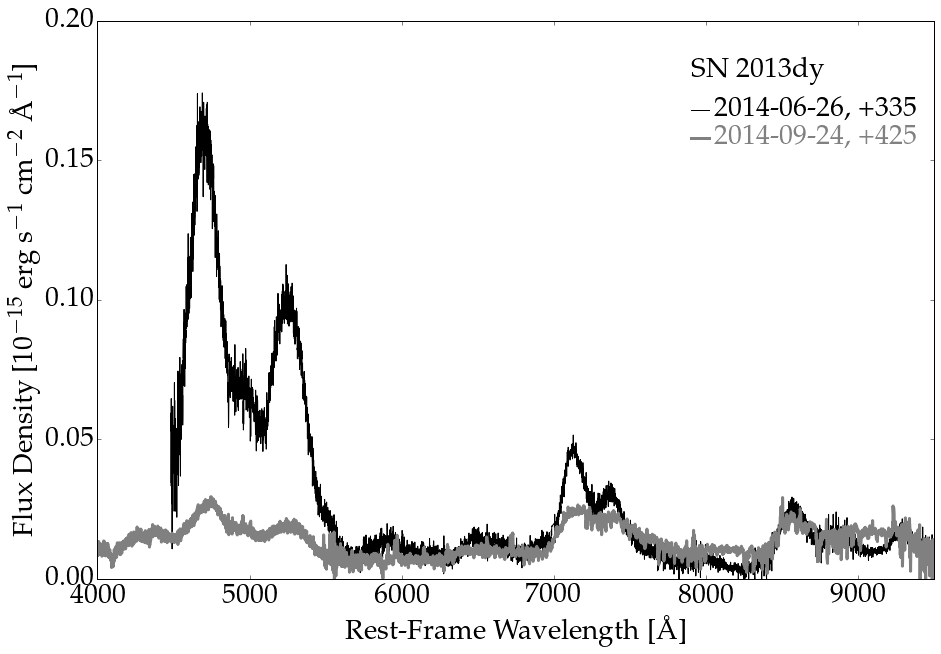}
\caption{Nebular-phase spectra for our eight SNe\,Ia, dereshifted into the rest frame and corrected for line-of-sight Galactic reddening (Table \ref{tab:objects}), and flux scaled to the late-time photometry where possible (Table \ref{tab:GMOSphot}). For SN\,2012fr we include an inset to compare the iron complex of lines by flux scaling the later epoch to have the same integrated flux between the dots as the earlier epoch; vertical lines mark the peak wavelengths. \label{fig:allspec_individual}}
\end{center}
\end{figure*}

\begin{figure*}
\begin{center}
\includegraphics[width=15cm,trim={5cm 2.5cm 6cm 4.2cm},clip]{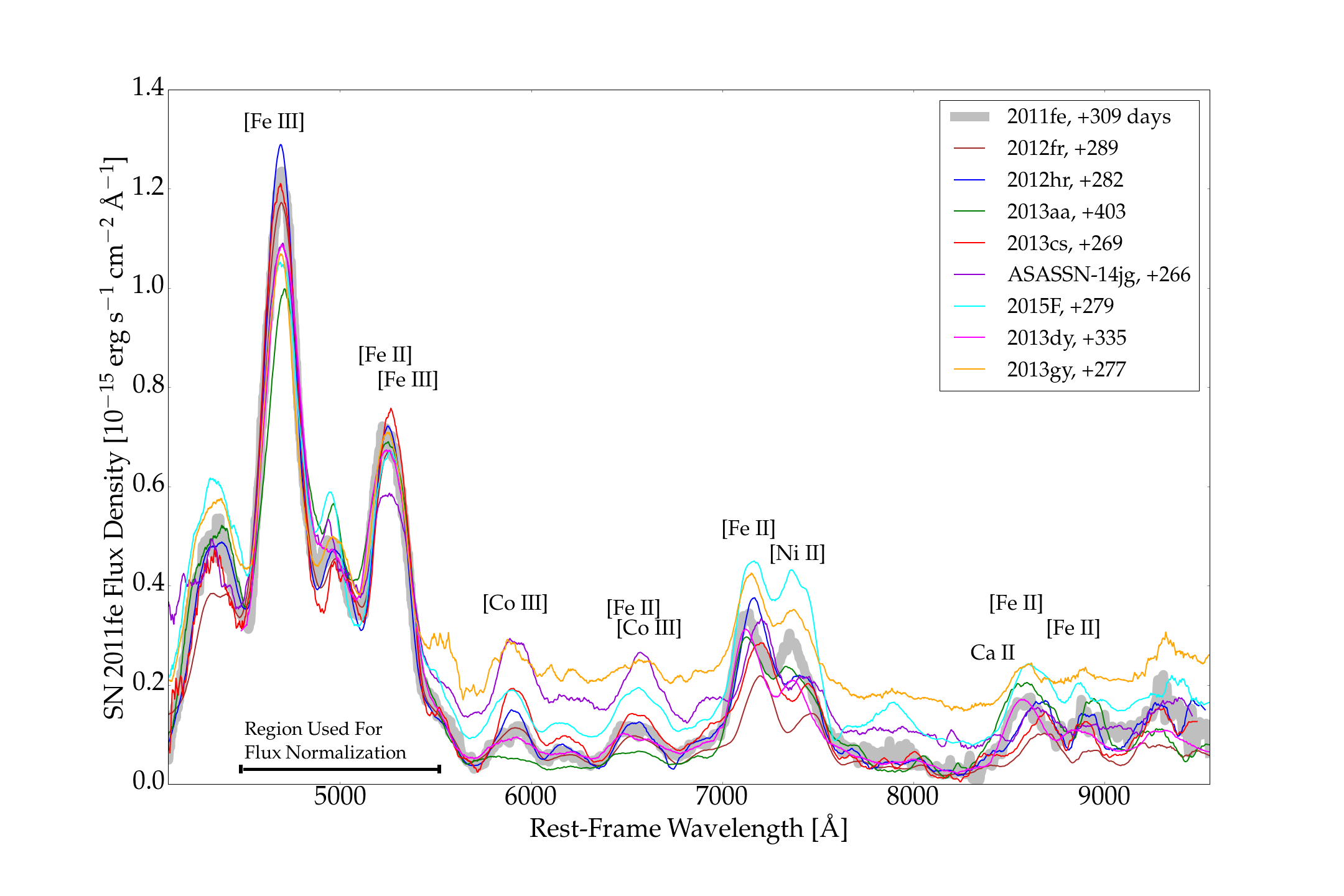}
\caption{Our eight nebular-phase SN\,Ia spectra from Gemini and Keck Observatories, along with a nebular-phase spectrum of the normal SN\,Ia 2011fe in grey for comparison. Wavelengths have been dereshifted into the rest frame, and the flux density has been corrected for line-of-sight Galactic reddening (Table \ref{tab:objects}), scaled to match the flux of SN\,2011fe between 4500 and 5500\,\AA, and smoothed with a Savitsky-Golay filter of width 50\,\AA\ and order 1. Note that the phases vary from +260 to +400\,days after peak brightness. Emission-feature labels represent the dominant species from modeled lines. \label{fig:allspec_comp}}
\end{center}
\end{figure*}

We begin our analysis with a visual comparison of the nebular-phase spectra in our sample that are listed in Tables \ref{tab:specobs} and \ref{tab:keckspecobs}. In Figure \ref{fig:allspec_individual} we plot the reduced and calibrated spectra for each object individually, with two epochs for each of SN\,2012fr (top left) and SN\,2013dy (bottom right) plotted to show the spectral evolution at late times. In Figure \ref{fig:allspec_comp} we show one rest-frame nebular-phase spectrum for each of the eight SNe\,Ia on a single plot to facilitate a comparison and highlight the diversity in nebular features. We also plot the $+309$\,day spectrum of the prototypically normal SN\,Ia 2011fe; this spectrum was obtained with the Kast Spectrograph at Lick Observatory, and previously published by \citep{2015MNRAS.446.2073G}. In particular, we note that while the lines attributed to [\ion{Fe}{III}] $\lambda4700$ and [\ion{Co}{III}] $\lambda 5900$ are similar in shape and velocity for most of our sample (although less similar in flux), the most diversity is seen in the red-side blend of [\ion{Fe}{II}] and [\ion{Ni}{II}].

For the nebular-phase emission lines we measure three parameters: the line velocity (i.e., from the central wavelength), the FWHM, and the integrated flux. These three parameters are connected to the physical state of the ejecta in the following ways. 
(1) The emission line for an expanding symmetric sphere of SN\,Ia ejecta material should be centred at zero velocity in the rest frame. Any offset is indicative of a net flow of the material toward or away from the observer, which indicates asymmetry in the nucleosynthetic products (e.g., \citealt{2010Natur.466...82M}).
(2) The FWHM represents the total distribution in velocity of the material along the line of sight, which is related to the density profile of the material at the time of explosion and its kinetic energy (e.g., \citealt{2015MNRAS.450.2631M}). 
(3) The integrated flux is a measure of the total amount of energy escaping the nebula, and is related to the synthesised mass of $^{56}$Ni and the physical state of the nebula (e.g., the temperature, ionisation state, positron trapping efficiency; \citealt{2015MNRAS.454.3816C}).

We use these data as an opportunity to compare two main methods for calculating emission-line parameters: fitting a Gaussian function and direct measurement. These two methods have been used by (for example) \cite{2013MNRAS.430.1030S}, \cite{2015MNRAS.446.2073G}, and \cite{2015MNRAS.454.3816C}. For the Gaussian fits, we follow the same procedure as did \cite{2015MNRAS.446.2073G}: we subtract a pseudocontinuum created by joining the local minima on each side of the emission line, and perform a least-squares fit of a Gaussian function (with two components for the blended [\ion{Fe}{II}] and [\ion{Ni}{II}] lines) to the non-normalised, flux-calibrated spectra and bootstrap our parameter uncertainties. For the direct measurements, we first subtract the pseudocontinuum and then smooth the spectrum (specifically, we use the {\sc IDL} routine {\sc POLY\_SMOOTH} with the Savitzky-Golay filter option and a 100\,\AA\ bin size). We then calculate the line velocity from the wavelength of the pixel at peak flux, the FWHM from the difference in wavelength between the two pixels at which the flux value is half the peak flux, and the total line flux by analytically integrating over the line. We estimate the uncertainties in these line parameters with a bootstrap method: we randomly scatter the pixel flux values by an amount proportional to their uncertainty, recalculate the line parameters in 300 trials, and take the standard deviation of the results as the parameter's error value. For the rest-frame central wavelengths of the nebular emission lines, we use [\ion{Fe}{III}] $\lambda4701$, [\ion{Co}{III}] $\lambda5891$, [\ion{Fe}{II}] $\lambda7155$, and [\ion{Ni}{II}] $\lambda7378$.

\begin{table*}
\begin{tabular}{llllccc}
\hline
\hline
SN Name & UT Date & Phase  & Line & Velocity & FWHM & Flux \\
 & (YYYYMMDD) & (days) & & (km\,s$^{-1}$) & (km\,s$^{-1}$) &  ($10^{-15}$\,erg\,s$^{-1}$\,cm$^{-2}$) \\
 \hline
SN\,2012fr & 20130829 & +289 & [Fe~III] & $-794\pm21$ & $12057\pm31$ & $92.2\pm0.7$ \\
               &                   &     & [Co~III] & $1267\pm51$ &$10192\pm75$ &$7.9\pm0.2$ \\
               &                   &     & [Fe~II] & $1927\pm23$ &$6998\pm21$ &$14.7\pm0.2$ \\
               &                   &     & [Ni~II] & $3437\pm36$ &$6667\pm40$ &$9.0\pm0.2$ \\
               & 20140102 & +415 & [Fe~III] & $-1788\pm76$ &$14488\pm93$ &$11.5\pm0.3$ \\
               &                   &     & [Co~III] & \ldots & \ldots & \ldots \\
               &                   &     & [Fe~II] & $1770\pm159$ &$9312\pm185$ &$4.3\pm0.3$ \\
               &                   &     & [Ni~II] & $3381\pm164$ &$7350\pm221$ &$2.8\pm0.3$ \\
SN\,2012hr & 20131006 & +283 & [Fe~III] & $-890\pm24$ &$10493\pm31$ &$15.6\pm0.2$ \\
               &                   &     & [Co~III] & $792\pm77$ &$8388\pm100$ &$1.7\pm0.1$ \\
               &                   &     & [Fe~II] & $395\pm61$ &$7205\pm83$ &$4.7\pm0.1$ \\
               &                   &     & [Ni~II] & $1325\pm124$ &$9233\pm140$ &$3.3\pm0.2$ \\
SN\,2013aa & 20140327 & +400 & [Fe~III] & $227\pm127$ &$11494\pm156$ &$33.6\pm1.7$ \\
               &                   &     & [Co~III] & $-3112\pm1314$ &$19764\pm2473$ &$2.5\pm0.9$ \\
               &                   &     & [Fe~II] & $-1591\pm70$ &$6376\pm83$ &$7.9\pm0.4$ \\
               &                   &     & [Ni~II] & $-1323\pm172$ &$12225\pm179$ &$12.4\pm0.5$ \\
SN\,2013cs & 20140211 & +265 & [Fe~III] & $-1106\pm204$ &$12182\pm250$ &$22.2\pm1.7$ \\
               &                   &     & [CoIII] & $1004\pm192$ &$8589\pm251$ &$2.6\pm0.3$ \\
               &                   &     & [FeII] & $2097\pm110$ &$7577\pm165$ &$3.5\pm0.2$ \\
               &                   &     & [NiII] & $3185\pm139$ &$6667\pm181$ &$2.3\pm0.2$ \\
ASASSN-14jg & 20150725 & +267 & [Fe~III] & $-655\pm161$ &$12307\pm218$ &$10.3\pm0.6$ \\
               &                   &     & [Co~III] & $1024\pm275$ &$9992\pm375$ &$2.1\pm0.3$ \\
               &                   &     & [Fe~II] & $1981\pm177$ &$6584\pm207$ &$2.3\pm0.2$ \\
               &                   &     & [Ni~II] & $3034\pm374$ &$8112\pm520$ &$1.5\pm0.3$ \\
SN\,2015F & 20151230 & +280 & [Fe~III] & $-540\pm114$ &$10994\pm156$ &$15.9\pm0.7$ \\
               &                   &     & [Co~III] & \ldots &$9491\pm150$ &$2.2\pm0.1$ \\
               &                   &     & [Fe~II] & $-111\pm81$ &$7453\pm83$ &$6.9\pm0.2$ \\
               &                   &     & [Ni~II] & $454\pm94$ &$9593\pm100$ &$8.7\pm0.3$ \\
SN\,2013dy & 20140626 & +335 & [Fe~III] & $-452\pm107$ &$12119\pm156$ &$33.8\pm1.4$ \\
               &                   &     & [Co~III] &  &$13633\pm819$ &$2.4\pm0.5$ \\
               &                   &     & [Fe~II] & $-1151\pm75$ &$6625\pm83$ &$7.3\pm0.3$ \\
               &                   &     & [Ni~II] & $-635\pm148$ &$7751\pm220$ &$4.5\pm0.4$ \\
               & 20140924 & +425 & [Fe~III] & $1429\pm158$ &$12806\pm218$ &$3.5\pm0.2$ \\
               &                   &     & [Co~III] & \ldots & \ldots & \ldots \\
               &                   &     & [Fe~II] & $-1422\pm180$ &$8610\pm206$ &$3.8\pm0.4$ \\
               &                   &     & [Ni~II] & \ldots &$13656\pm436$ &$5.6\pm0.5$ \\
SN\,2013gy & 20140924 & +275 & [Fe~III] & $-1026\pm64$ &$10242\pm94$ &$3.7\pm0.1$ \\
               &                   &     & [Co\,III] & $-1221\pm555$ &$9691\pm925$ &$0.6\pm0.2$ \\
               &                   &     & [Fe\,II] & $-433\pm141$ &$6542\pm166$ &$1.0\pm0.1$ \\
               &                   &     & [Ni\,II] & $227\pm220$ &$9072\pm340$ &$1.0\pm0.1$ \\
\hline
\end{tabular}
\caption{Parameters of the nebular-phase emission lines, obtained by fitting Gaussian functions.}
\label{tab:gaussian_parameters}
\end{table*}

\begin{table*}
\begin{tabular}{llllccc}
\hline
\hline
SN Name & UT Date & Phase  & Line & Velocity & FWHM & Flux \\
 & (YYYYMMDD) & (days) & & (km\,s$^{-1}$) & (km\,s$^{-1}$) &  ($10^{-15}$\,erg\,s$^{-1}$\,cm$^{-2}$) \\
 \hline
SN\,2012fr & 20130829 & +289 & [Fe~III] & $-924\pm15$ & $12244\pm31$ &$87.7\pm2.06$ \\
               &                   &     & [Co~III] & $1094\pm26$ &$10242\pm25$ &$7.33\pm0.393$ \\
               & 20140102 & +415 & [Fe~III] & $-1878\pm39$ &$14363\pm31$ &$10.3\pm0.417$ \\
               &                   &     & [Co~III] & $264\pm319$ &$15221\pm322$ &$0.444\pm0.171$ \\
SN\,2012hr & 20131006 & +283 & [Fe~III] & $-951\pm19$ &$10555\pm31$ &$14.9\pm0.425$ \\
               &                   &     & [Co~III] & $805\pm43$ &$8137\pm50$ &$1.64\pm0.169$ \\
SN\,2013aa & 20140327 & +400 & [Fe~III] & $272\pm67$ &$11306\pm94$ &$32.4\pm2.47$ \\
               &                   &     & [Co~III] & $-1544\pm411$ &$16014\pm470$ &$1.51\pm0.557$ \\
SN\,2013cs & 20140211 & +265 & [Fe~III] & $-1021\pm106$ &$11932\pm125$ &$20.3\pm2.46$ \\
               &                   &     & [Co~III] & $1291\pm65$ &$7987\pm75$ &$2.27\pm0.217$ \\
ASASSN-14jg & 20150725 & +267 & [Fe~III] & $-655\pm68$ &$12119\pm94$ &$9.56\pm0.729$ \\
               &                   &     & [Co~III] & $1207\pm90$ &$8940\pm100$ &$1.76\pm0.318$ \\
SN\,2015F & 20151230 & +280 & [Fe~III] & $-636\pm51$ &$10931\pm63$ &$14.6\pm1.24$ \\
               &                   &     & [Co~III] & $-162\pm61$ &$9240\pm75$ &$2.00\pm0.268$ \\
SN\,2013dy & 20140626 & +335 & [Fe~III] & $-476\pm65$ &$12119\pm62$ &$31.2\pm3.32$ \\
               &                   &     & [Co~III] & $-146\pm230$ &$10842\pm275$ &$1.48\pm0.745$ \\
               & 20140924 & +425 & [Fe~III] & $961\pm61$ &$11744\pm63$ &$3.11\pm0.439$ \\
SN\,2013gy & 20140924 & +275 & [Fe~III] & $-1247\pm33$ &$10179\pm31$ &$3.53\pm0.327$ \\
               &                   &     & [Co~III] & $-1573\pm313$ &$7535\pm352$ &$0.395\pm0.293$ \\
\hline
\end{tabular}
\caption{Parameters of the nebular-phase emission lines, obtained from direct measurements.}
\label{tab:direct_parameters}
\end{table*}

\begin{figure}
\begin{center}
\includegraphics[width=8.5cm, trim={1.5cm 0.5cm 0.8cm 1cm},clip]{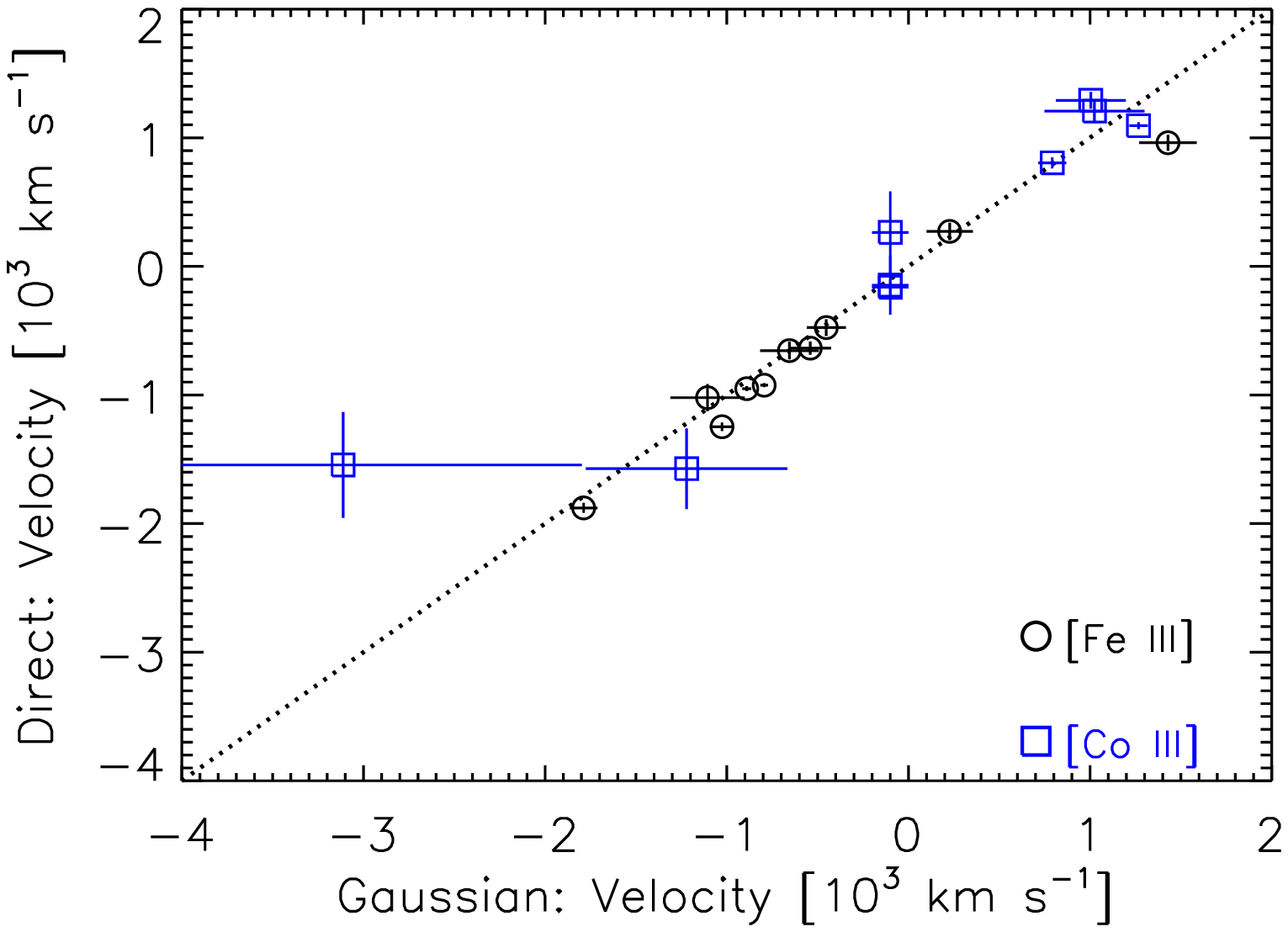}
\includegraphics[width=8.5cm, trim={1.3cm 0.5cm 0.5cm 0.8cm},clip]{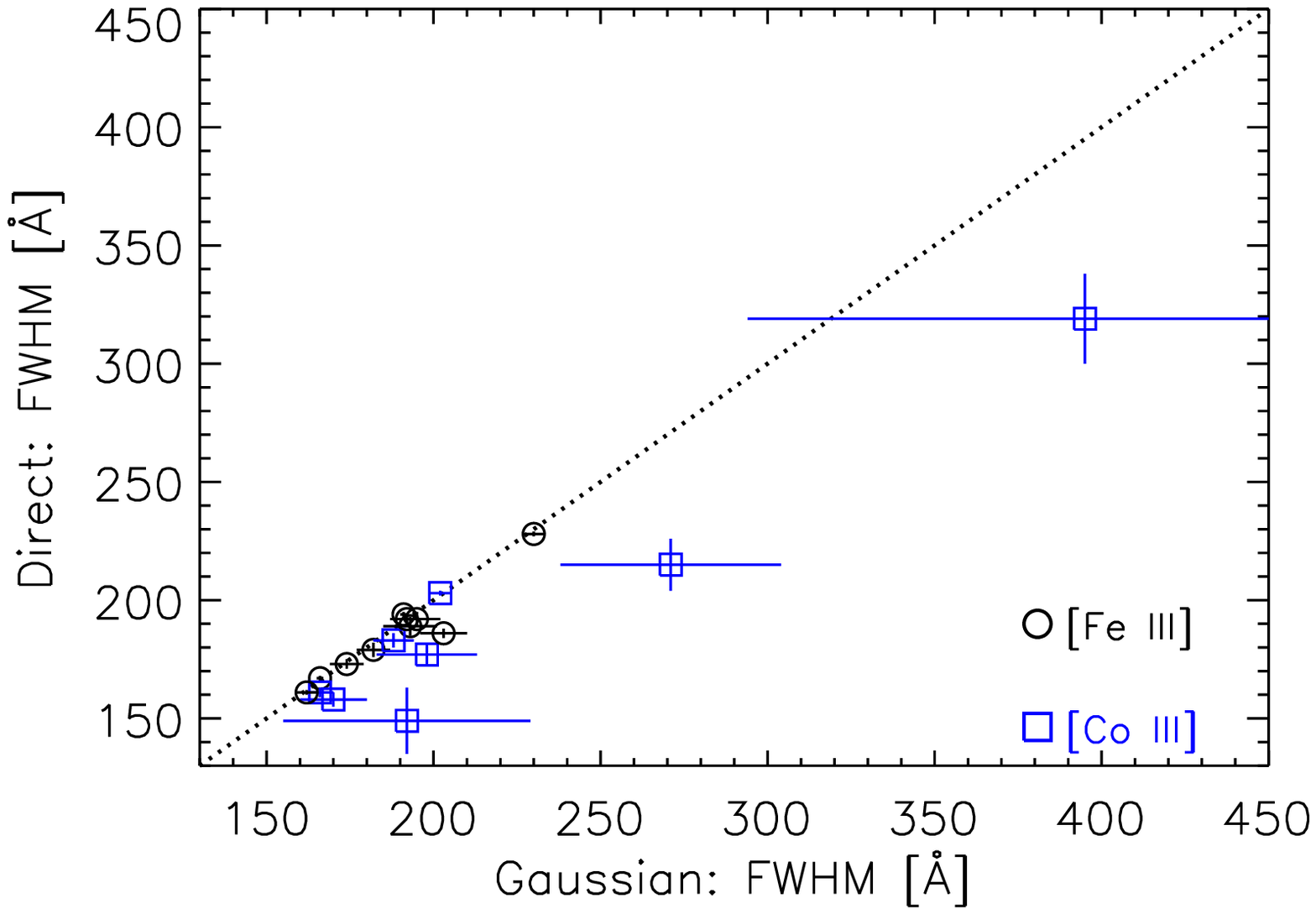}
\includegraphics[width=8.5cm, trim={1.2cm 0.4cm 0.5cm 0.7cm},clip]{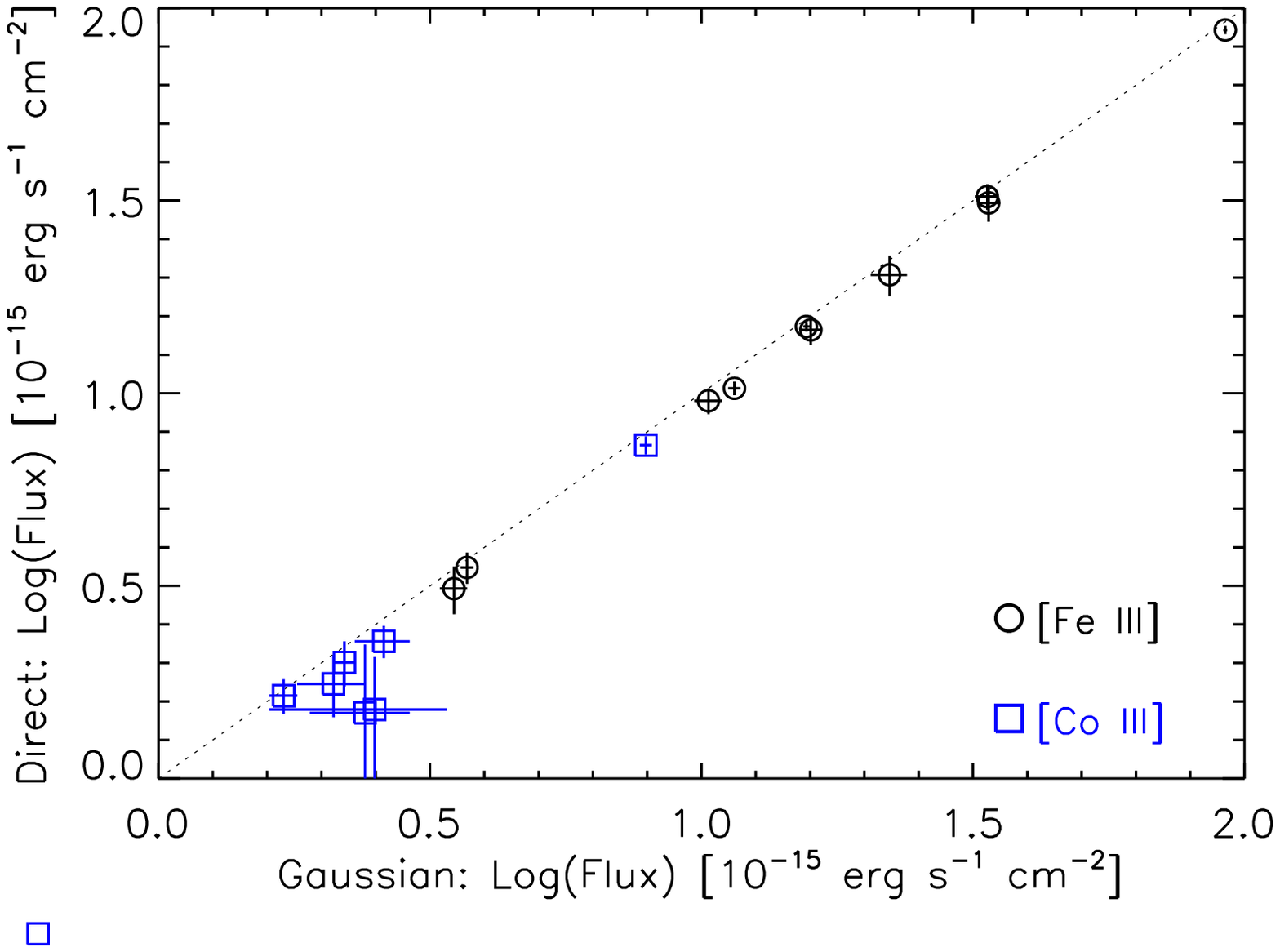}
\caption{Comparison of the emission-line parameters determined from the direct (ordinate) and Gaussian fit (abscissa) methods, for velocity (top), FWHM (centre), and integrated flux (bottom). The blue and black points represent the [Fe~III] and [Co~III] lines, respectively. The [Fe~II] and [Ni~II] lines are not included because they are blended; direct measures of their parameters are not possible. We can see that the two methods yield similar line parameters that are within the uncertainties. For lines which are not actually a Gaussian, the direct method produces significantly more precise results. \label{fig:compare}}
\end{center}
\end{figure}

The resulting line parameters and their uncertainties for the Gaussian fits and the direct measurements are presented in Tables \ref{tab:gaussian_parameters} and \ref{tab:direct_parameters}, respectively. Note that the blended [\ion{Fe}{II}] and [\ion{Ni}{II}] lines' parameters can only be measured with the Gaussian fitting method. The two methods are compared for each measurement in Figure \ref{fig:compare}. From these plots we take away three main points: (1) the values deviate the most when the uncertainties are large, (2) the direct method leads to more precise measurements in general, and (3) there is a small systematic offset in the Gaussian-fit integrated fluxes. This offset was also remarked on by \cite{2015MNRAS.454.3816C} (i.e., their Figure 3), and is caused by deviations in the line profile from Gaussianity. In addition to the measurement errors of these emission features, there are systematics introduced by line-blending with weaker features from other species. For example, Figure 3 of \cite{2017arXiv170406275B} demonstrates how models of nebular spectra indicate that the [\ion{Fe}{III}] $\lambda \approx 4700$\,\AA\ and the [\ion{Co}{III}] $\lambda \approx 5900$\,\AA\ lines are reasonably pure, but that the [\ion{Fe}{II}] and [\ion{Ni}{II}] blend can be contaminated by [\ion{Ca}{II}], especially in SNe\,Ia that synthesise relatively little stable nucleosynthetic products. Fortunately, it appears that strong contamination from calcium would also cause the two emission lines to be quite heavily blended into almost a single feature, whereas we see a clear double peak in all our spectra (Fig. \ref{fig:allspec_comp}). To fully quantify the impact of line blending requires detailed modeling which we consider beyond the scope of this analysis.

In the following sections we analyze our observations for each of the SNe\,Ia in our sample individually, and put them in context with other existing published observations where available. One of these contexts that we will frequently refer to is the off-centre explosion model of \cite{2010Natur.466...82M}, proposed to explain the observed relation that HVG (LVG) SNe\,Ia also exhibit nebular emission lines of [\ion{Fe}{II}] and [\ion{Ni}{II}] that are redshifted (blueshifted or unshifted). In this model, an off-centre ignition shifts the deflagration burning either toward or away from the observer's line of sight, imparting a blueshift or a redshift to the nebular emission lines of the nucleosynthetic products of the deflagration: [\ion{Fe}{II}] and [\ion{Ni}{II}]. After the deflagration phase, the detonation proceeds throughout the rest of the material. \cite{2010Natur.466...82M} find that on the side opposite to the off-centre ignition, the burned material has a shallower density distribution, and suggest that a SN\,Ia observed from this side will exhibit a higher \ion{Si}{II} absorption-line velocity and velocity gradient. This is how an ignition offset away from the observer along the line of sight will result in an HVG SN\,Ia with redshifted nebular lines, and vice versa. In the following sections we will refer regularly to this correlation between the \ion{Si}{II} velocity gradient and the nebular emission-line velocity, and discuss whether each of our SNe\,Ia exhibits characteristics that are consistent with the model of \cite{2010Natur.466...82M}.

\subsection{SN\,2012fr}\label{ssec:sn2012fr}

SN\,2012fr was discovered on October 27, 2012 (UT dates are used throughout this paper), two weeks before it reached peak brightness \citep{2012CBET.3275....1K}. It was classified as a young Type Ia SN with a spectrum obtained on October 28, 2012 \citep{2012CBET.3275....2C}. Spectra obtained three and five days later exhibited a \ion{Si}{II} $\lambda 6355$ feature with the main photospheric component at $\sim 14,000$\,km\,s$^{-1}$ and a detached, high-velocity component at $\sim 19,000$\,km\,s$^{-1}$ (high-velocity \ion{Ca}{II}~H\&K was also seen; \citealt{2012ATel.4538....1H}). SN\,2012fr was well offset from its host, NGC\,1365, a face-on Sb-type galaxy at a distance of $D=18.07\pm2.7$\,Mpc ($z=0.005457$; \citealt{1996ApJ...463...60B})\footnote{The redshift-independent summary statistic of 50 distances in the literature, from the NASA Extragalactic Database, NED: \url{https://ned.ipac.caltech.edu/}.}. As SN\,2012fr was nearby, radio observations were triggered and \cite{2012ATel.4567....1K} report a nondetection at 5.8\,GHz with the Very Large Array on October 30, 2012 with an upper limit of $18$ $\rm \mu Jy$. Pre-explosion archival {\it Hubble Space Telescope} imaging existed for the host galaxy; \cite{2012ATel.4535....1G} report a nondetection at the location of the SN, and put an upper limit on the progenitor system brightness of $M_V\gtrsim-5.9$\,mag, ruling out a supergiant companion star.

Analyses of the evolution in the optical photometric, spectroscopic, and spectropolarimetric features of SN\,2012fr are presented in the literature. \cite{2013ApJ...770...29C} give a dense time series of optical spectra, showing that the high-velocity component of \ion{Si}{II} $\lambda 6355$ fades by 5\,days before peak brightness, and that the photospheric component is narrow and remains at $v\approx 12,000$\,km\,s$^{-1}$ after peak brightness, classifying SN\,2012fr as an LVG SN\,Ia. They also point out that while the overluminous peak brightness and slow decline of SN\,2012fr resemble those of SN\,Ia 1991T \citep{1992ApJ...384L..15F, 1992AJ....103.1632P}, spectroscopically SN\,2012fr sits on the border of SN\,Ia subclasses (normal {\it vs.} high-velocity, and shallow-silicon {\it vs.} core-normal in the scheme of \citealt{2006PASP..118..560B}). 

\cite{2013MNRAS.433L..20M} present spectropolarimetry around peak brightness and use the \ion{Ca}{II} near-infrared triplet to determine that the polarisation angle of the high-velocity component is oriented at 90\degree\ from that of the low-velocity components of \ion{Ca}{II} and \ion{Si}{II}, and that the overall continuum polarisation is low. They suggest that their observations support a delayed-detonation model over a merger model for SN\,2012fr, and furthermore predict that the iron-group nebular lines will be blueshifted (based on the model of \citealt{2010Natur.466...82M}). 

\cite{2014AJ....148....1Z} present an analysis of optical and ultraviolet (UV) photometry covering the full light curve out to $\sim120$\,days, plus a time series of optical spectroscopy to $+71$\,days after peak brightness. They evaluate the bolometric light curve of SN\,2012fr and derive the mass of $^{56}$Ni synthesised in the explosion, $M_{\rm Ni} = 0.88 \pm 0.08$\,M$_{\odot}$, which they point out is smaller than that of SN\,1991T but larger than that of the paragon of normal SNe\,Ia, SN\,2011fe. 

We find that the nebular-phase spectra of SN\,2012fr deviate from expectations in two ways. First, although it was securely classified as an LVG SN\,Ia based on its $\dot{v}_{\rm Si~II} \approx 0$ km\,s$^{-1}$, the nebular lines of [\ion{Fe}{II}] and [\ion{Ni}{II}] exhibit a redshift of $\sim 2500$\,km\,s$^{-1}$ instead of the blueshift generally seen for LVG SNe\,Ia \citep{2010Natur.466...82M}. We speculate that these redshifted nebular lines may be related to the relatively high \ion{Si}{II} absorption-line velocity of SN\,2012fr, $v_{\rm Si~II}({\rm peak}) \approx 12,000$ km\,s$^{-1}$\,d$^{-1}$, which is more consistent with an HVG-like SN\,Ia and would correlate with redshifted nebular emission lines in the offset ignition model. Second, the velocity of the [\ion{Fe}{III}] $\lambda4700$ line is typically observed to progress redward over time in the nebular phase, but in SN\,2012fr it appears to be significantly bluer, by $\sim -1400$\,km\,s$^{-1}$, at $+415$\,days than at $+289$\,days (see the inset plot in Figure \ref{fig:allspec_individual}). Although the later epoch's [\ion{Fe}{III}] $\lambda4700$ line has a broader, flatter shape and lower S/N than the earlier epoch --- which could cause a bluer peak wavelength to be reported from our fits --- we find that this blueward progression is not an artifact. In the top-left panel of Figure \ref{fig:allspec_individual}, the inset plot compares the iron complex for the two epochs, clearly showing how their blue-side edge broadens while the red-side edge remains at the same wavelength, which is mainly driving the shift in the peak of the line. The evolution in iron line velocities at late times are discussed further in Section \ref{ssec:FeIII}.

\subsection{SN\,2012hr}

SN\,2012hr was discovered in an image taken on December 16, 2012 by the Backyard Observatory Supernova Search (BOSS; \citealt{2012CBET.3346....1D}) in the host galaxy ESO\,121-G026 (also known as PGC\,18880; host offset 2\arcsec\ W and 94\arcsec\ N), a barred spiral galaxy located at redshift $z = 0.007562$ \citep{2004AJ....128...16K}. The Carnegie Supernova Project obtained a near-infrared spectrum of SN\,2012hr on December 20, 2012 and classified it as a SN\,Ia approximately a week before peak brightness, exhibiting \ion{Mg}{II} absorption at $12,200$\,km\,s$^{-1}$ \citep{2012ATel.4663....1M}. SN\,2012hr was also slightly overluminous (Table \ref{tab:lcfit}), similar to SN\,2012fr. The similarity of SN\,2012hr to SN\,2012fr extends to late times, with similar shapes and velocities for their [\ion{Fe}{III}] and [\ion{Co}{III}] lines, as seen in Figure \ref{fig:allspec_comp} and Table \ref{tab:gaussian_parameters}. 

As shown in Table \ref{tab:vSiII} and discussed in Section \ref{ssec:early}, we directly measure the \ion{Si}{II} velocity gradient of SN\,2012hr to be $\dot{v}_{\rm Si~II} = 120 \pm 80$ km\,s$^{-1}$ and classify it as an HVG SN\,Ia. However, since the large uncertainty is consistent with LVG SNe\,Ia ($\dot{v}_{\rm Si~II} < 70$ km\,s$^{-1}$), and its \ion{Si}{II} velocity ($v_{\rm Si~II}({\rm peak}) \approx 11,500$ km\,s$^{-1}$\,d$^{-1}$) is lower than typically seen for HVG SNe\,Ia ($\sim 12,000$ km\,s$^{-1}$\,d$^{-1}$), we consider this classification insecure and label it as ``HVG?". In this way, it is dissimilar to SN\,2012fr. In our nebular spectra of SN\,2012hr we find that the [\ion{Fe}{II}] and [\ion{Ni}{II}] lines are redshifted, as expected for an HVG object, but by only a small amount ($<1000$\,km\,s$^{-1}$): this is smaller than most of the HVG SNe\,Ia in the sample of \cite{2010Natur.466...82M}, but still consistent with the lower boundary (i.e., SN\,2002er in their Figure 2). The relationship between photospheric and nebular line velocities is further discussed in Section \ref{ssec:asym}.

\subsection{SN\,2013aa}

SN\,2013aa was discovered by the Backyard Observatory Supernova Search (BOSS) on February 13, 2013 \citep{2013CBET.3416....1P}. SN\,2013aa is 74\arcsec\ W and 180\arcsec\ S from the core of NGC\,5643, a barred spiral galaxy at $D \approx 16.9$\,Mpc \citep{Tully1988}. Follow-up observations by the Las Cumbres Observatory on February 15, 2013 spectroscopically confirmed SN\,2013aa as a SN\,Ia near maximum light \citep{2013ATel.4817....1P}. We securely classify SN\,2013aa as an LVG SN\,Ia based on a direct measurement of $\dot{v}_{\rm Si~II} = 50 \pm 10$ km\,s$^{-1}$ (Table \ref{tab:vSiII}), and also find that it was overluminous at peak brightness (Table \ref{tab:lcfit}).

Our nebular-phase spectrum of SN\,2013aa is at a much later phase (+400\,days) than the other SNe\,Ia in our sample. This was possible because it is one of the most nearby objects, and thus brighter in apparent magnitude. At +400\,days we see that the [\ion{Fe}{III}] $\lambda4700$ line is redshifted, similar to SNe\,Ia at late times in general (see Section \ref{ssec:FeIII} for further discussion). The emission lines of [\ion{Fe}{II}] and [\ion{Ni}{II}] are both blueshifted, consistent with expectations for an LVG SN\,Ia under the model of \cite{2010Natur.466...82M}. We also find that [\ion{Co}{III}] $\lambda5900$ has a broader, flatter profile in SN\,2013aa than some of our other spectra, but that it is similar to our spectra of SN\,2012fr and SN\,2013dy at comparably late phases, $>1$\,yr after maximum light. An interpretation of [\ion{Co}{III}] line-profile shapes at late nebular phases is discussed in Section \ref{ssec:collision}.

\subsection{SN\,2013cs}

SN\,2013cs was first discovered by the La Silla Quest Survey (as LSQ13aiz) on May 12, 2013 in the host galaxy ESO\,576-G017 \citep{2013ATel.5067....1W}, and independently discovered on May 14, 2013 by the Catalina Real-time Transient Survey in an unfiltered Siding Spring Survey Image \citep{2013CBET.3533....1A}. SN\,2013cs was classified as an SN\,Ia by Yamanaka and collaborators using an optical spectrum obtained on May 15, 2013 \citep{2013CBET.3533....1A}. This spectrum, at a phase of $-10$\,days (i.e., before peak brightness), exhibited \ion{Si}{II} $\lambda6355$ at a high velocity of $15,000$\,km\,s$^{-1}$ and the rarely observed \ion{C}{II} $\lambda 6580$ absorption feature. At peak brightness the \ion{Si}{II} velocity had decreased to $12,500$\,km\,s$^{-1}$, but this is still quite high, similar to that exhibited by HVG SNe\,Ia. We do not have any photospheric-phase spectra in the two weeks post-maximum, so we cannot confirm that it belongs in the HVG class, and have instead given it the insecure designation of ``HVG?" in Table \ref{tab:vSiII}.

Our $+269$\,day nebular-phase spectrum of SN\,2013cs is similar to that of SN\,2012fr and SN\,2012hr, as seen in Figure \ref{fig:allspec_comp}. The emission lines of [\ion{Fe}{II}] and [\ion{Ni}{II}] in SN\,2013cs are redshifted by an average of $v_{\rm neb}\approx2600$\,km\,s$^{-1}$. This is a much higher velocity than in SN\,2012hr, our only other ``HVG?"-classified SN\,Ia, and consistent with the expectation for an HVG object in the asymmetric explosion model of \cite{2010Natur.466...82M} (discussed further in Section \ref{ssec:asym}).

\subsection{ASASSN-14jg}

ASASSN-14jg was discovered by the All Sky Automated Survey for SuperNovae (ASASSN) on October 29, 2014 using the double 14~cm Cassius telescope in Cerro Tololo, Chile \cite{2014ATel.6637....1H}. It is located 8.5\arcsec\ north and 12.5\arcsec\ east from the core of PGC\,128348, a spiral galaxy at $D = 59.6$\,Mpc. ASASSN-14jg was classified as an SN\,Ia with a FLOYDS spectrum obtained on the Faulkes Telescope South \cite{2014ATel.6661....1A}. A spectrum obtained on November 6, 2014 by the ANU WiFeS SuperNovA Program (AWSNAP) showed no local H$\alpha$ emission but did exhibit the \ion{Na}{I} $\lambda\lambda$5890, 5896  absorption feature at a velocity of $8.9 \pm 15.3$\,km\,s$^{-1}$ with respect to the SN host galaxy --- i.e., most likely associated with the host-galaxy interstellar medium and not potential evidence of circumstellar material \citep{2016arXiv160708526C}. As shown in Tables \ref{tab:lcfit}, ASASSN-14jg exhibited a normal peak brightness and decline rate. We could not directly measure the velocity gradient, and instead designate ASASSN-14jg as an ``LVG?" SN\,Ia based on its low velocity of photospheric \ion{Si}{II}, $v_{\rm Si~II}({\rm peak}) \approx 10,200$ km\,s$^{-1}$\,d$^{-1}$, as described in Section \ref{ssec:early} (see also Table \ref{tab:vSiII}). 

Our nebular-phase spectrum of ASASSN-14jg at $+266$\,days appears to exhibit a relatively higher continuum flux on the red side, as seen in Figure \ref{fig:allspec_comp} where all the spectra are scaled to have equal integrated blue-side fluxes. However, the conditions at Gemini South were poor, with variable clouds on the night of the observation, making it difficult to properly flux calibrate the separately obtained blue-side and red-side spectra. Further evidence of this is the large (0.2--0.3\,mag) uncertainty in our Gemini South photometry from that night (Table \ref{tab:GMOSphot}). The nebular-phase emission lines of [\ion{Fe}{II}] and [\ion{Ni}{II}] are significantly redshifted, with an average $v_{\rm neb}\approx2500$\,km\,s$^{-1}$. This is inconsistent with expectations for LVG-like SNe\,Ia in the model of \cite{2010Natur.466...82M}, which typically exhibit blueshifted nebular emission. This is discussed in further detail in Section \ref{ssec:asym}.

\subsection{SN\,2015F}

SN\,2015F was discovered on March 9, 2015 in host galaxy NGC\,2422 \citep{2015CBET.4081....1M}, a peculiar, weakly barred spiral galaxy at $D = 20.5$\,Mpc. SN\,2015F is offset by 2.5\arcmin\ from the host-galaxy core \citep{2016arXiv160501054G}. PESSTO obtained an optical follow-up spectrum on March 11, 2015 that confirmed SN\,2015F as an SN\,Ia, resembling the spectrum of SN\,2002bo  $\sim12$\,days before peak brightness. They further subclassify SN\,2015F as at the boundary of normal and SN\,1991bg-like \citep{2015CBET.4081....1M,2015ATel.7209....1F}\footnote{See \citet{1992AJ....104.1543F, 1993AJ....105..301L} for a description of SN\,1991bg.} SN\,2015F reached peak brightness on March 25, 2015 with a subluminous maximum $B$-band magnitude of $-18.4 \pm 0.8$ \citep{2016arXiv160501054G}, and exhibited a decline rate of $\Delta m_B(15) \approx 1.2$\,mag. While this is the largest $\Delta m_B(15)$ value in our sample of SNe, and SN\,2015F does lie just below the \citet{1993ApJ...413L.105P} relation (e.g., Fig. 1 of \citealt{2013FrPhy...8..116H}), the decline is significantly slower than that of a typical SN\,1991bg-like SN\,Ia ($\Delta m_B(15) \approx 1.9$\,mag) and more similar to that of a normal SN\,Ia.

\cite{2017MNRAS.464.4476C} find that the early-time spectra of SN\,2015F exhibit strong \ion{Ca}{II} high-velocity features and an absorption feature at $\sim6800$\,\AA, which has been noted in several SN\,1991bg-like SNe\,Ia. They attribute this absorption feature to either photospheric \ion{Al}{II} or unburned carbon material in the outer layers of the SN ejecta. Aluminum has not been observed in other normal SNe\,Ia, but it has been reported in peculiar SNe like SN\,2010X \cite{2010ApJ...723L..98K}.

\cite{2015ApJS..221...22I} present first-light photometry of SN\,2015F from their daily-cadenced survey and identify extremely early-time emission occurring $\sim1.5$\,days before the light curve begins to rise in earnest. Specifically, they find two 2$\sigma$ detections out of 38 observations, on days $-2$ and $-3$. Statistically, one should observe $0.05\times38 = 1.9$ detections at 2$\sigma$ in such a dataset, so these observations are not inconsistent with a nondetection. Despite this, \citet{2015ApJS..221...22I} interpret this as the signature of ejecta interacting with a nondegenerate companion star of at least $R>0.1$ $\rm R_{\odot}$. In this progenitor scenario we may see swept-up hydrogen radiating at late times, but given the underlying host-galaxy emission for SN\,2015F, that is unfortunately impossible to confirm with our data. 

Returning to the suggestion that SN\,2015F's subluminous peak brightness might make it similar to the SN\,1991bg-like class, we consider whether SN\,2015F is SN\,1991bg-like at late times. In our $+279$\,day optical nebular spectrum of SN\,2015F, we observe that the line velocities and widths are similar to those of other normal nebular-phase SNe\,Ia of our sample (i.e., SN\,2013aa excluded). In this way the nebular features of SN\,2015F are dissimilar to those of SN\,1991bg, the latter being dominated by narrow [\ion{Fe}{III}] at a similar phase \citep{1996MNRAS.283....1T}, which was suggested by \cite{2012MNRAS.424.2926M} to be from a high degree of ionisation in the nebula at late times. Our nebular-phase spectrum of SN\,2015F exhibits [\ion{Fe}{II}] and [\ion{Ni}{II}] emission lines that are nearly equal in peak flux, which is unique among our sample of SNe\,Ia. As [\ion{Fe}{II}] represents a product of the decay of radioactive $^{56}$Ni and any Ni at late times must have been formed stably, this observation qualitatively indicates that SN\,2015F produced a relatively lower ratio of radioactive to stable nucleosynthetic products compared to other SNe\,Ia. This may be an indication of a higher than average metallicity for the progenitor star of SN\,2015F. In this scenario, the additional neutrons from the higher abundance of heavy elements could lead to more stable nuclear burning (e.g., as investigated by \citealt{2015MNRAS.446.2073G}). As a final note, the nearly equal peak flux of the [\ion{Fe}{II}] and [\ion{Ni}{II}] emission lines of SN\,2015F is similar to the case of SN\,1991bg, although the latter were more blended. \cite{2012MNRAS.424.2926M} show that model explosions with a low central density, a low mass of radioactive $^{56}$Ni (both of which can affect the ionisation state at late times), and/or an additional boost from \ion{Ca}{II} at 7200\,\AA\ can all generate simulated spectra that match those of SN\,1991bg in this respect, but it is difficult to isolate the underlying physical cause.

For SN\,2015F we could not directly measure the velocity gradient, and instead designate it as an ``LVG?" SN\,Ia because it exhibited a low photospheric \ion{Si}{II} velocity, $v_{\rm Si~II}({\rm peak}) \approx 10,100$ km\,s$^{-1}$\,d$^{-1}$, as described in Section \ref{ssec:early} and listed in Table \ref{tab:vSiII}. The average nebular-phase emission-line velocities of [\ion{Fe}{II}] and [\ion{Ni}{II}] is $v_{\rm neb} \approx 340 \pm 120 $ km\,s$^{-1}$ (from Table \ref{tab:gaussian_parameters}), slightly redshifted yet still consistent with the upper bound for LVG SNe\,Ia from \cite{2010Natur.466...82M} (i.e., $<1000$ km\,s$^{-1}$, from their Figure 2). 

\subsection{SN\,2013dy}

SN\,2013dy was discovered by the Lick Observatory Supernova Search \citep[LOSS;][]{2001ASPC..246..121F, 2011MNRAS.412.1419L} with the 0.76~m Katzman Automatic Imaging Telescope (KAIT) in an image taken on July 10, 2013 \citep{2013CBET.3588....1C}. It was detected only 2.4\,hr after first light (equivalent to the explosion time unless there is a dark phase, which might last many hours), making it the earliest-observed SN\,Ia \citep{2013ApJ...778L..15Z}. SN\,2013dy is located in the host galaxy NGC\,7250, an irregular/diffuse spiral galaxy at $D \approx 12.4$\,Mpc (this is a mean of redshift- and SN-independent distance measurements\footnote{\url{ned.ipac.caltech.edu}} from \citealt{2009AJ....138..323T}, \citealt{2011A&A...532A.104N}, and \citealt{2013AJ....146...86T}). SN\,2013dy is offset $2.3\arcsec$ west and $26.4\arcsec$ north of its host galaxy's core ($\sim1.76$ kpc). It was spectroscopically classified as a normal SN\,Ia with a strong \ion{C}{II} absorption feature at early times \citep{2013ApJ...778L..15Z}.

\cite{2015MNRAS.452.4307P} present a thorough analysis of SN\,2013dy with UV-optical-infrared photometry and spectroscopy up to $+500$\,days post-explosion. With early-time photospheric-phase spectra, they securely classify SN\,2013dy as a member of the LVG subclass. They find that the light curve declines slowly, like that of SN\,1991T, but existing redshift-independent distances (based on the Tully-Fisher relation) indicate that it was not overluminous like SN\,1991T, even after significant host extinction ($A_{V,{\rm host}}=0.64$\,mag) is accounted for \citep{2015MNRAS.452.4307P}. However, they also show that the distance modulus derived directly from the light curve of SN\,2013dy is significantly larger. In Table \ref{tab:lcfit}, we quote the intrinsic peak $B$-band magnitude of $M_{\rm peak}(B) = -19.66$, based on the SN-derived distance and not the Tully-Fisher relation. This is at the bright end of the \citet{1993ApJ...413L.105P} relation (e.g., Figure 13 of \cite{2012AJ....143..126B}).

With their late-time optical spectra of SN\,2013dy, \cite{2015MNRAS.452.4307P} show how the [\ion{Fe}{III}] line velocity evolves redward from $\sim-2700$ $\rm km\ s^{-1}$ at $+100$ days to $+1350$ $\rm km\ s^{-1}$ at $+400$ days. This is in agreement with our own measurement of $-476$ $\rm km\ s^{-1}$ at $+335$ days. In the asymmetric model of \cite{2010Natur.466...82M}, this blueshifted nebular emission velocity is consistent with expectations for SN\,2013dy, as it is a member of the LVG subclass. \cite{2015MNRAS.452.4307P} find that SN\,2013dy exhibits weaker [\ion{Fe}{II}] and [\ion{Ni}{II}] than the paragon SN\,2011fe, as is also seen in our Figure \ref{fig:allspec_comp}. This leads them to infer a larger mass of $^{56}$Ni for SN\,2013dy than for SN\,2011fe, which they note is consistent with the broader light curve (see paragraph above).

\subsection{SN\,2013gy}

SN\,2013gy was discovered by LOSS with KAIT on December 6, 2013 in the host galaxy NGC\,1418 \cite{2013ATel.5637....1Z}, offset 35.08\arcsec\ from the nucleus \citep{2016arXiv160501054G}. NGC\,1418 is a barred spiral galaxy at $D \approx 66$\,Mpc (a mean of redshift-independent distances\footnote{\url{ned.ipac.caltech.edu}} from \citealt{2007A&A...465...71T}, \citealt{2009ApJS..182..474S}, and \citealt{2013AJ....146...86T}). SN\,2013gy was classified as a normal SN\,Ia with an optical spectrum obtained on December 7, 2013 \citep{2013ATel.5638....1T}, consistent with its normal light curve. We did not have the necessary post-peak spectra to directly measure the velocity gradient for SN\,2013gy, and instead designate it as an ``LVG?" SN\,Ia because it exhibited a low photospheric \ion{Si}{II} velocity, $v_{\rm Si~II}({\rm peak}) \approx 10,700$ km\,s$^{-1}$\,d$^{-1}$, as described in Section \ref{ssec:early} and listed in Table \ref{tab:vSiII}.

In our $+275$\,day optical nebular spectrum, SN\,2013dy appears normal and similar to the others, except that like ASASSN-14jg, it seems to have a relatively large red-side flux. In spectra of faint objects that have a flux gradient near the 5600\,\AA\ dichroic, joining the blue and red sides of an LRIS spectrum can be problematic, and in Figure \ref{fig:allspec_individual} we see substantial noise at 5600\,\AA. Co-epochal photometry (as we obtained for some objects with GMOS) can help resolve such issues, but we did not obtain LRIS photometry of SN\,2013gy. However, this is not a serious issue, because our analysis does not rely on an absolute flux calibration; our measurements of line widths, velocities, and continuum-subtracted fluxes should be unaffected. We find that the lines of [\ion{Fe}{II}] and [\ion{Ni}{II}] are slightly blueshifted, with an average $v_{\rm neb} \approx -100$\,km\,s$^{-1}$, consistent with expectations of LVG-like SNe\,Ia in the model of \cite{2010Natur.466...82M}, as discussed in Section \ref{ssec:asym}.

\section{Discussion}\label{sec:disc}

As in the analysis above, we focus our discussion on the attributes of our individual events and provide physical interpretations for their characteristics with respect to SN\,Ia progenitor scenarios and explosion mechanisms. When appropriate, we compare the distribution of derived parameters for our eight events to those of larger published datasets, but otherwise we find that our sample size is too small to provide new statistical constraints on correlations involving nebular-phase spectra of SNe\,Ia (e.g., the relation between emission-line redshift/blueshift and the photospheric \ion{Si}{II} velocity gradient). We hope that future analyses of the large, communally available samples of nebular-phase SN\,Ia spectra will incorporate our data and the derived properties, but 
such a study is beyond the scope of this paper. We now discuss the explosion asymmetry (Section \ref{ssec:asym}), limits on hydrogen in the system (Section \ref{ssec:Ha}), and the redward progression of the $\sim4700$\,\AA\ feature (Section \ref{ssec:FeIII}).

\subsection{Explosion Asymmetry}\label{ssec:asym}

\begin{figure}
\begin{center}
\includegraphics[width=8.5cm]{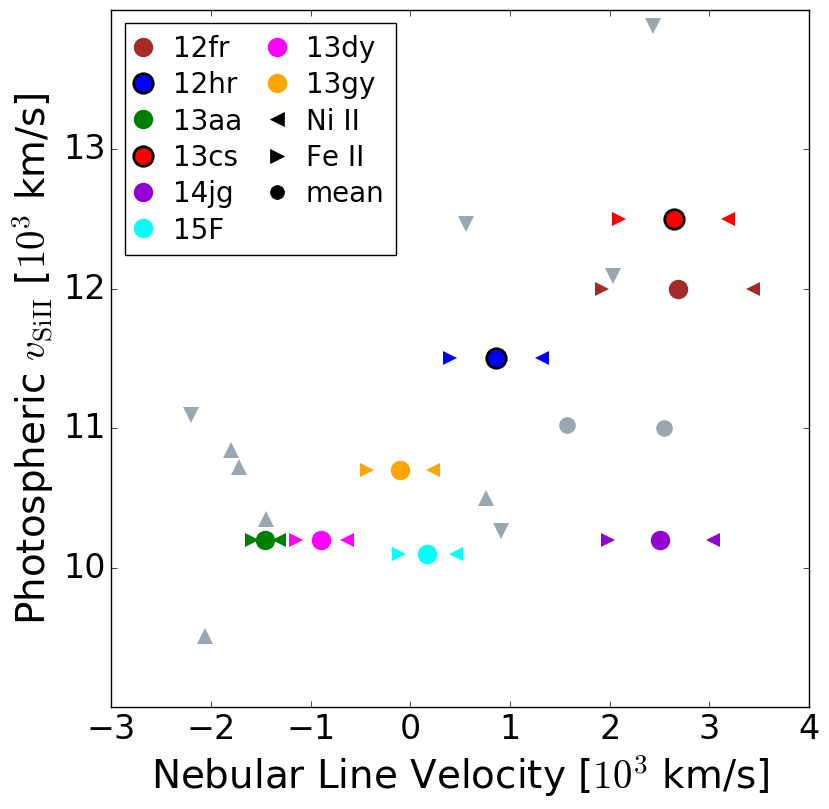}
\includegraphics[width=8.5cm]{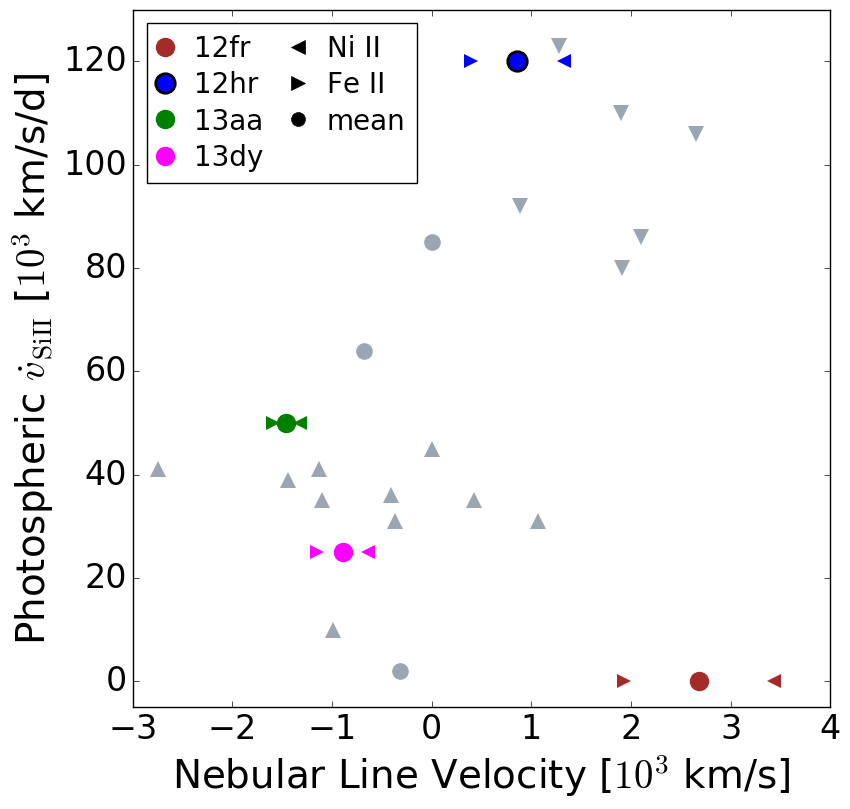}
\caption{{\it Top:} For each of the SNe\,Ia in our sample of nebular-phase spectra we plot the velocity of the photospheric-phase \ion{Si}{II} absorption line at peak brightness as a function of the velocity of the [\ion{Fe}{II}] and [\ion{Ni}{II}] emission lines (triangles) and their average velocity (circles), from our Table \ref{tab:gaussian_parameters}. Each SN\,Ia is marked with a different colour as described by the legend, and the two that we tentatively associate with the HVG group, SN\,2012hr and SN\,2013cs, are outlined in black. The grey points represent the sample of nebular SNe\,Ia from \protect\cite{2013MNRAS.430.1030S}, where upward triangles are LVG, downward triangles are HVG, and circles represent other classifications of SNe\,Ia. {\it Bottom:} For the four SNe\,Ia in our sample with a direct measurement of the \ion{Si}{II} velocity gradient, $\dot{v}_{\rm Si~II}$ (Section \ref{ssec:early}), we plot this as a function of the velocity of the [\ion{Fe}{II}] and [\ion{Ni}{II}] emission lines in the same manner as the top panel. The grey symbols represent the sample of nebular SNe\,Ia from \protect\cite{2010Natur.466...82M}, with the same symbol type convention.
\label{fig:vneb_vSiII}}
\end{center}
\end{figure}

As described briefly elsewhere in this paper, an off-centre ignition followed by an asymmetric SN\,Ia explosion has been proposed as the underlying physical cause of  an observed correlation between the photospheric-phase velocity gradient of \ion{Si}{II} and the nebular-phase velocity of the emission lines of the nucleosynthetic products \citep{2010Natur.466...82M}. In the model of Maeda et al., the outer layers of the ejecta on the same side as the explosion are compacted, and relatively denser than the outer ejecta layers on the opposite side. A SN\,Ia in which the explosion is offset {\it away} from the observer (on the far side along the line of sight) exhibits photospheric \ion{Si}{II} line velocities that evolve more rapidly because the photosphere recedes more quickly down into the slower layers of ejecta --- i.e., they belong to the HVG group. A SN\,Ia with this orientation with respect to the observer will also exhibit redshifted nebular-phase emission lines from the nucleosynthetic products. This is reversed for SNe\,Ia which are observed with an offset explosion on the near side, {\it toward} the observer: an LVG SN\,Ia with blueshifted nebular-phase emission lines. A near-side offset also predicts a narrower light curve and fainter peak luminosity, and so has been proposed as a contributing factor to the width vs. luminosity correlation \citep{2011MNRAS.413.3075M}.

\cite{2010Natur.466...82M} initially demonstrated the relation between the photospheric \ion{Si}{II} velocity gradient and the nebular emission line velocity in a small sample of 17 SNe\,Ia, and this trend was confirmed by \cite{2013MNRAS.430.1030S} with a larger sample. Unfortunately, our small sample has only one tentative HVG SN\,Ia with a direct measurement of the \ion{Si}{II} velocity gradient: SN\,2012hr exhibits $\dot{v}_{\rm Si~II} = -120\pm80$\,km\,s$^{-1}$\,d$^{-1}$, but its large uncertainty makes it consistent with the LVG group. As discussed in Section \ref{ssec:early}, HVG SNe\,Ia typically also exhibit higher-velocity photospheric \ion{Si}{II} near maximum light, and by using this method we also tentatively identify SN\,2013cs as a potential HVG SN\,Ia. 

In the top panel of Figure \ref{fig:vneb_vSiII} we plot the photospheric-phase velocity of \ion{Si}{II} versus the nebular-phase velocity of the [\ion{Fe}{II}] and [\ion{Ni}{II}] emission lines for all our SNe\,Ia, along with the sample from \cite{2013MNRAS.430.1030S} for comparison. In the bottom panel of Figure  \ref{fig:vneb_vSiII} we plot the \ion{Si}{II} velocity gradient vs. the nebular line velocity for the four SNe\,Ia with directly measured $\dot{v}_{\rm Si~II}$, along with the sample from \cite{2010Natur.466...82M} for comparison.

Our goal here is not to confirm the relation (or the proposed physical model) --- our sample is too small to make an impact in this regard, even if we had direct measures of $\dot{v}_{\rm Si~II}$ for all objects --- but to put our sample of SNe\,Ia in context. We instead remark that we see the same general structure in this diagram as did \cite{2010Natur.466...82M} and \cite{2013MNRAS.430.1030S}: the tentative HVG-group SNe\,Ia exhibit redshifted nebular-velocity emission lines associated with the nucleosynthetic products. The two notable exceptions here are SN\,2012fr and ASASSN-14jg, which had redshifted nebular emission lines but were classified as LVG, but there are several caveats to mention. SN\,2012fr was classified as an LVG SN\,Ia with directly measured $\dot{v}_{\rm Si~II}$, but as we can see in the top panel of Figure \ref{fig:vneb_vSiII}, the velocity of its photospheric \ion{Si}{II} absorption line was similar to that of other HVG SNe\,Ia. Based on the clearly and significantly redshifted nebular lines, SN\,2012fr remains a candidate for an off-centre ignition oriented away from the observer, but would be a special case in which the associated photospheric-phase \ion{Si}{II} velocity gradient, $\dot{v}_{\rm Si~II}$, does not evolve as expected by the model of \cite{2010Natur.466...82M}. If, as suggested by \cite{2013ApJ...770...29C}, the ejecta of SN\,2012fr were stratified and had only a thin layer of partial-burning products at $v>12000$ $\rm km\ s^{-1}$, then it was simply impossible to observe the effect of a fast-receding photosphere using the \ion{Si}{II} absorption line in this case. ASASSN-14jg was classified as an LVG SN\,Ia based on its low photospheric \ion{Si}{II} velocity, $v_{\rm Si~II} = 10,200$\,km\,s$^{-1}$, but the earliest spectrum was obtained 12\,days after peak brightness. If it had a $\dot{v}_{\rm Si~II} = 150$\,km\,s$^{-1}$\,d$^{-1}$, then $v_{\rm Si~II}({\rm peak}) \approx 12,000$\,km\,s$^{-1}$. Ultimately, we find that all of our observations are potentially consistent with an off-centre ignition model, although some deviations from the relation between nebular line velocity and $\dot{v}_{\rm Si~II}$ from \cite{2010Natur.466...82M} are required.

\subsubsection{Collisional SN\,Ia Candidates}\label{ssec:collision}

\begin{figure*}
\begin{center}
\includegraphics[width=15cm,trim={4cm 2.5cm 6cm 4.2cm},clip]{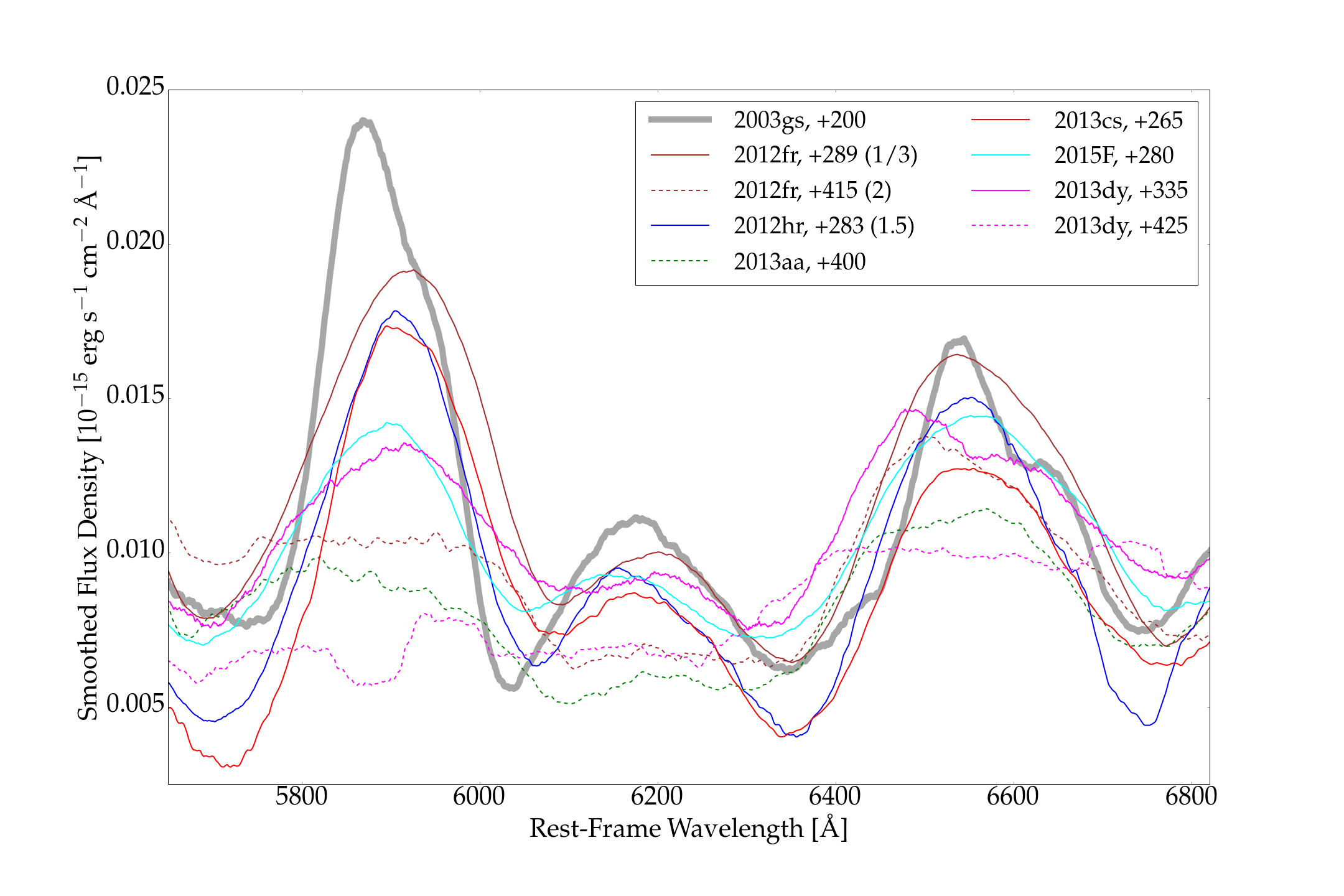}
\caption{Eight of our nebular-phase spectra of SNe\,Ia in the region of [\ion{Co}{III}] $\lambda5900$ and $\lambda 6200$ and the [\ion{Fe}{II}] and [\ion{Co}{III}] blended feature at 6600\,\AA. We include a nebular-phase spectrum of SN\,2003gs at $+200$\,days post-maximum light (previously published by \protect\citealt{2013MNRAS.430.1030S}) for comparison, because SN\,2003gs is one of three collisional candidates identified to exhibit two-component emission features by \protect\cite{2015MNRAS.454L..61D}. All spectra are corrected to rest-frame redshifts and have their apparent flux densities smoothed using a Savitsky-Golay filter of polynomial order 1 with a window size of 75\,\AA\ for clarity. For some SNe, the fluxes have been scaled in order to aid the comparison and the scale factors are shown in parentheses in the legend. Nebular spectra obtained on or later than $+400$ after peak brightness are shown with a dashed line to distinguish them from the earlier nebular-phase spectra. \label{fig:CoIII}}
\end{center}
\end{figure*}

Asymmetric explosions are also predicted by the double white dwarf collision scenario (e.g., \citealt{2009ApJ...705L.128R,2009MNRAS.399L.156R}) in which two carbon-oxygen white dwarf stars collide. They could be two single stars in an environment of high local stellar density, or a binary system with altered trajectories from (for example) tidal interactions or the Kozai mechanism \citep{2011ApJ...741...82T}. A tell-tale signature of the collisional model is predicted at late times if the collision is aligned with the observer's line of sight: double-peaked nebular-phase emission lines from the two regions of nucleosynthetic products with distinct velocities, as identified in [\ion{Co}{III}] by \cite{2015MNRAS.454L..61D}. Depending on the viewing angle, a collision-triggered SN\,Ia could also have nebular-phase emission lines that appear relatively normal or broadened.

We look for collisional candidates in our sample of SNe\,Ia by examining the profiles of their [\ion{Co}{III}] features. In Figure \ref{fig:CoIII} we plot our nebular-phase spectra in the region of [\ion{Co}{III}] along with the nebular-phase spectrum of SN\,2003gs \citep{2013MNRAS.430.1030S}, which is one of the 3 SNe\,Ia exhibiting double-peaked [\ion{Co}{III}] identified by \cite{2015MNRAS.454L..61D} as a collisional candidate. SNe\,Ia ASASSN-14jg and 2013gy are omitted from this plot because their spectra suffer from an unfilled chip gap and poor S/N in this region, respectively. We do not find any clear indication of two velocity components for the [\ion{Co}{III}] $\lambda5900$ line of SNe\,2012fr (at $289$\,days), 2012hr, 2013cs, or 2015F, but we do find this line has broadened and deviated from a Gaussian shape in SNe\,2012fr (at $415$\,days), 2013aa, and 2013dy. The former group's spectra are all at $<300$\,days, while the latter's were obtained $>300$\,days after peak brightness. Although broadened features can be created by multiple velocity components, we find it more likely that the shape of this feature is simply common in nebular-phase spectra at and beyond 1\,yr after peak brightness. A similar observation of a broad and flattened 5900\,\AA\ feature was made for prototypically normal SN\,2011fe at $>1$\,yr by \cite{2015MNRAS.454.1948G}. They postulated that the observed emission near 5900\,\AA\ may instead be from \ion{Na}{I}~D, which is predicted by models to dominate during later nebular phases ($>1$\,yr). 

None of our SNe\,Ia fall into the low-luminosity, fast-declining subgroup in which \cite{2015MNRAS.454L..61D} find all their collisional candidates (i.e., $\Delta m_{15}(B)>1.3$\,mag). In fact, SNe\,2012fr, 2013dy, and 2013aa were the three most intrinsically luminous of our sample. While it may not necessarily be that all collisions produce underluminous SNe\,Ia --- the existing models are able to predict a wide range of $\rm ^{56}Ni$ masses \citep{2013ApJ...778L..37K} --- there is another potential underlying physical cause of broad line profiles in SNe\,Ia. Overluminous SNe\,Ia create a higher mass of radioactive nickel, have more energy to impart to their ejecta, and can thus exhibit broader FWHM in their nebular-phase emission lines. However, in this scenario the line profiles would still retain a Gaussian shape.

Although we find that some of our nebular spectra have broadened, non-Gaussian emission features near [\ion{Co}{III}] $\lambda5900$, none exhibits the clear double-peaked profiles seen by \cite{2015MNRAS.454L..61D}, and ultimately we cannot confirm them as collisional candidates.

\subsection{Hydrogen at Nebular Phases}\label{ssec:Ha}

\begin{figure}
\begin{center}
\includegraphics[width=8.5cm]{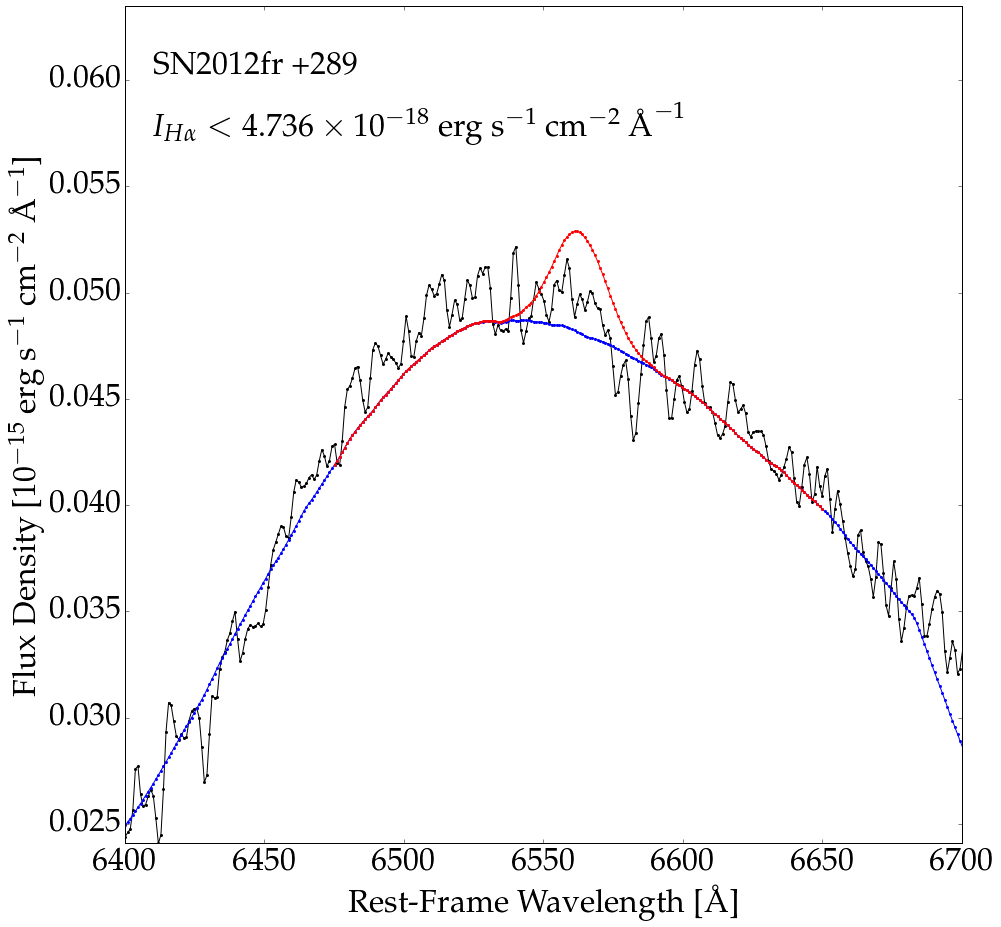}
\includegraphics[width=8.5cm]{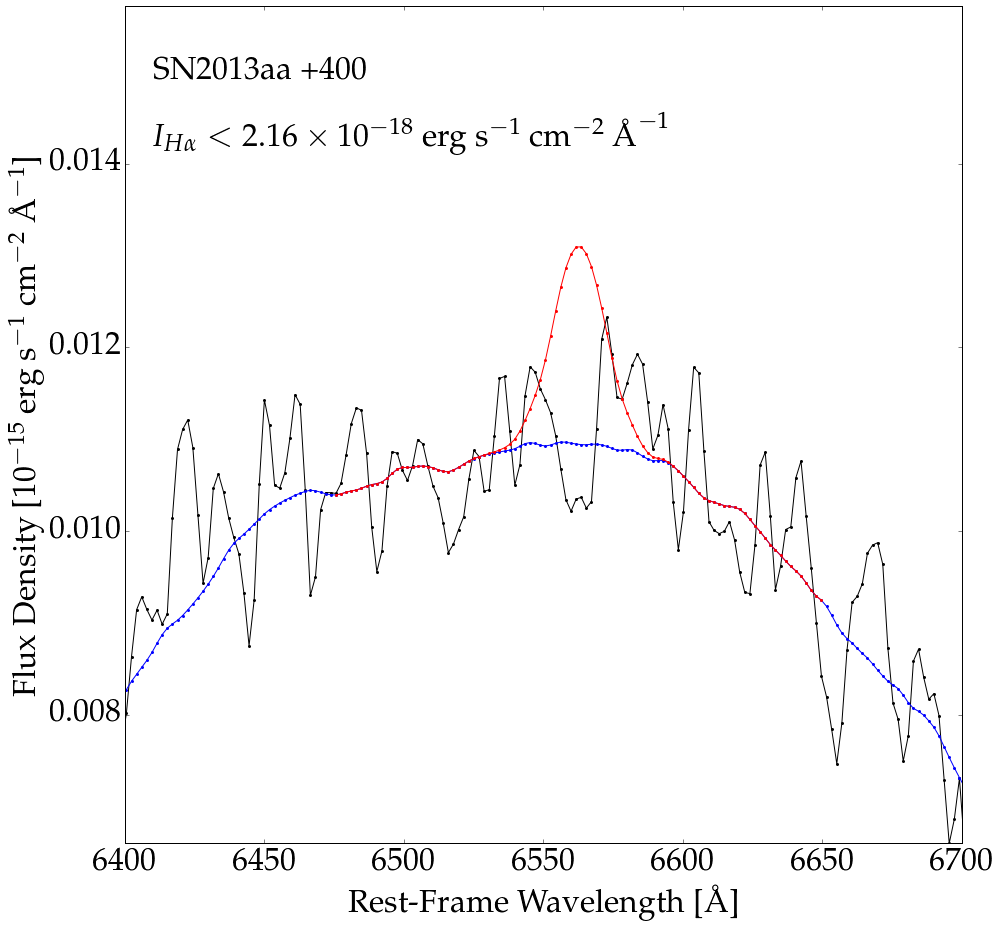}
\caption{Constraints on narrow H$\alpha$ emission in the two SNe in our sample for which the spectral S/N is sufficiently good, and the local environment sufficiently clean, to attempt this measurement. Black represents the flux spectrum and blue represents the smoothed flux spectrum; the root-mean-square (RMS) noise is calculated from their difference in the wavelength region shown. The red represents the upper limit on H$\alpha$ as a Gaussian emission line with FWHM $=22$\,\AA\ and a peak of 3 times the RMS, and the peak intensity $I_{{\rm H}\alpha}$ is printed in the upper-left corner. \label{fig:halpha}}
\end{center}
\end{figure}

In the single-degenerate model for SNe\,Ia, mass is accreted onto the primary white dwarf from a main-sequence or red-giant companion star. In this scenario, unaccreted hydrogen from the companion could be swept-up by the SN\,Ia ejecta and create an observable emission line with a velocity $v\approx -1000$\,km\,s$^{-1}$ and a FWHM $\approx 22$~\AA\ (e.g., \citealt{2013ApJ...778..121L}, Figure 9). Recently, \cite{2016MNRAS.457.3254M} presented the results of a statistical search for hydrogen in 11 SNe\,Ia and found a tentative detection in one of them. Despite our selection criteria that the objects we observe be well separated from the host galaxy, a visual inspection of the 2D spectra in the region of H$\alpha$ reveals that extended H$\alpha$ emission from the host intersects with the trace of the SN in four of our eight objects.

The four SNe\,Ia in apparently ``clean'' environments are SNe\,2012fr, 2012hr, 2013aa, and 2013gy. However, for SNe\,2012hr the residuals from sky-line subtraction in the 1D spectrum in the vicinity of H$\alpha$ cause the S/N to be too poor to constrain the presence of H$\alpha$ emission. Also, we find that SN\,2013gy is projected onto a $z \approx 0.52$ background galaxy exhibiting narrow emission lines of H$\alpha$, H$\beta$, H$\gamma$, and [\ion{O}{III}], and that the H$\beta$ line appears near to where we would look for H$\alpha$ at SN\,2013gy's redshift. In Figure \ref{fig:halpha}, we show our spectra in the region of H$\alpha$ for SNe\,2012fr and 2013aa, in which we find no discernible emission line. 

To determine an upper limit on the H$\alpha$ emission-line flux in our spectra, we follow an analysis similar to that of \cite{2015MNRAS.454.1948G}, which in turn is based on the methodology of \cite{2001PASP..113..920L}, \cite{2007ApJ...670.1275L}, and \cite{2013ApJ...762L...5S}, and on the models of \cite{2000ApJS..128..615M} and \cite{2005A&A...443..649M}; see also searches for hydrogen in near-infrared SN\,Ia spectra by \cite{2016ApJ...822L..16S}. To summarise briefly, the maximum size of an undetected emission line is modeled as a Gaussian with FWHM $\approx 22$~\AA\ and a peak intensity 3 times the RMS flux error in the region. For the flux uncertainties we use the residual between the data and the local continuum, which we estimate by applying to the data a least-squares polynomial smoothing filter with a wide window. In Figure \ref{fig:halpha} we draw this maximum undetected H$\alpha$ emission line onto our spectra, and write the peak intensity in the upper-left corner of the plot. We use the distances in Table \ref{tab:objects} to find that the intrinsic upper limit on the H$\alpha$ intensity is $\sim$ (1--2) $\times10^{35}$\,erg\,s$^{-1}$\,\AA$^{-1}$ for SNe\,2012fr and 2013aa, which we convert to upper limits on the mass of hydrogen in the system of $<$ (1--2) $\times 10^{-2}$\,M$_{\odot}$ (following the same method as, say, \citealt{2015MNRAS.454.1948G}). Ultimately, we find that our limits on the mass of hydrogen in the systems of SN\,2012fr and SN\,2013aa are quite high, and that we cannot use them to constrain the progenitor scenarios. 

\subsection{The Redward Evolution of [\ion{Fe}{III}]}\label{ssec:FeIII}

\begin{figure}
\begin{center}
\includegraphics[width=8.5cm]{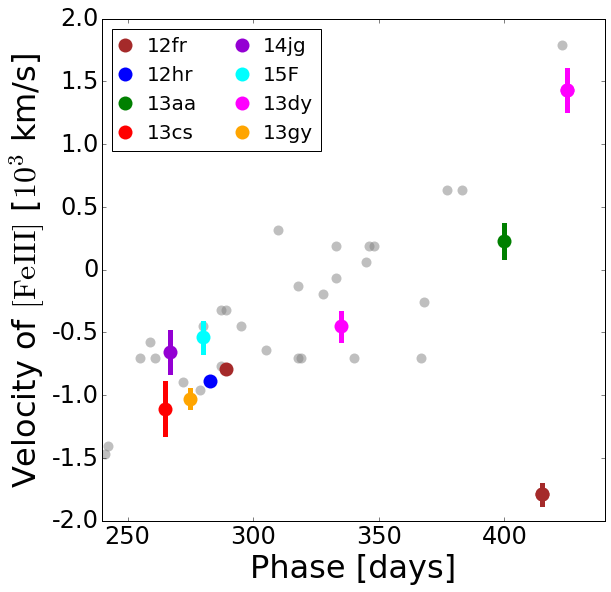}
\caption{For our sample of nebular-phase spectra, the velocity of the [\ion{Fe}{III}] emission line in each SN\,Ia is shown as a function of the phase in days past peak brightness, where the line velocities are determined by fitting a Gaussian function to the data (i.e., from Table \ref{tab:gaussian_parameters}). Each SN\,Ia is marked as a different colour as described by the legend. The grey circles in the background represent the dataset of \citet[their Table A1]{2016MNRAS.462..649B}. \label{fig:FeIII}}
\end{center}
\end{figure}

The continuous redward progression of the blueshifted peak of the [\ion{Fe}{III}] feature at $\lambda \approx 4700$\,\AA\ was first noted by \cite{2010ApJ...708.1703M} and confirmed in a larger sample of nebular-phase SN\,Ia spectra by \cite{2013MNRAS.430.1030S}. The latter find that the evolution in the velocity of [\ion{Fe}{III}] begins at $-4000$ to $-2000$\,km\,s$^{-1}$ at 100\,days after peak brightness, increases by 10--20\,km\,s$^{-1}\,d^{-1}$, and reaches $v\approx0$\,km\,s$^{-1}$ at $>200$\,days after peak brightness. They also find that the evolution appears to continue, with the line becoming redshifted. 

Recently, the evolution in the nebular-phase emission lines was explored using models for SN\,2011fe by \cite{2015ApJ...814L...2F}. They find that a change in the dominant species contributing to the line flux from \ion{Fe}{III} and \ion{Fe}{II} to \ion{Fe}{I} is the underlying cause of the apparent velocity shift, as the emission lines of \ion{Fe}{I} are just redward of \ion{Fe}{III} and \ion{Fe}{II}. Additionally, \cite{2016MNRAS.462..649B} combine spectral synthesis modeling with a sample of 160 nebular-phase spectra for 27 normal SNe\,Ia; they find that line blending, changes in opacity, and contributions from permitted iron lines may all participate in the redward evolution of the nebular-phase \ion{Fe} lines at $\lambda \approx 4700$\,\AA. 

All of our spectra were obtained $>200$\,days past maximum brightness, and for most of our objects we have only one epoch, so we cannot confirm or investigate the above findings in the same way. Instead, in Figure \ref{fig:FeIII} we plot the [\ion{Fe}{III}] $\lambda4700$ line velocity as a function of phase for the SNe\,Ia in our sample, and compare with the dataset of \cite{2016MNRAS.462..649B}. We see that all of our $<300$\,day spectra still exhibit a blueshifted [\ion{Fe}{III}] $\lambda4700$ line, and our sample does demonstrate the general trend that spectra at later phases exhibit higher-velocity \ion{Fe}{III} lines --- with the exception of SN\,2012fr at $>400$\,days. As presented in Section \ref{ssec:sn2012fr} and Figure \ref{fig:allspec_individual}, we find that the blue-side edges of the [\ion{Fe}{III}] $\lambda\lambda$4700, 5300 emission lines have moved to bluer wavelengths at the later epoch, while the red-side edges match fairly well: SN\,2012fr is a true outlier in terms of it's $400$ day [\ion{Fe}{III}] $\lambda$4700 line velocity. Given the work of \cite{2016MNRAS.462..649B} described in the preceding paragraph, we suspect that this is the result of a different opacity or ionisation level in the nebula of SN\,2012fr compared to other SNe\,Ia -- especially considering that SN\,2012fr was a more energetic explosion, forming a larger mass of $^{56}$Ni as described in Section \ref{ssec:sn2012fr}. \cite{2013ApJ...770...29C} and \cite{2013MNRAS.433L..20M} also discuss how the early-time observations of SN\,2012fr suggest the explosion may have been a delayed detonation or a double detonation, and the constraint of a surface layer of He could have affected the nebular-phase evolution. However, given that \cite{2015ApJ...814L...2F} find a myriad of underlying physical causes at play in generating the observed line positions in nebular spectra, and mindful of the risk of overinterpreting our analysis, we leave further evaluation to detailed models of the explosion of SN\,2012fr, which are beyond the scope of this work (see, e.g., the fine treatment of SN\,2011fe by \citealt{2015MNRAS.450.2631M}).

\section{Conclusions}\label{sec:conc}

We have presented new nebular-phase spectra from Gemini and Keck Observatories for eight nearby SNe\,Ia which were well observed at early times by the Las Cumbres Observatory. Data products for these SNe\,Ia were derived at early and late times, including light-curve parameters such as peak brightness and $B$-band decline rate, photospheric-phase spectral parameters such as the velocity of the \ion{Si}{II} absorption line, and nebular-phase emission-line parameters such as FWHM and velocity. We provided an analysis comprised of biographies for each of our SNe\,Ia as individual, unique events, remarking on particular characteristics and discussing instances where predictions about the late-time qualities from the early-time behaviour were or were not observed. For example, based on its photospheric-phase \ion{Si}{II} velocity, SN\,2012fr should have exhibited blueshifted nebular emission lines, but did not. 

We demonstrated the diversity in nebular-phase features of SNe\,Ia by comparing our eight spectra with each other and with the fiducial SN\,Ia 2011fe, and interpreted these features with respect to various physical models for SN\,Ia progenitor scenarios and explosion mechanisms. None of our SNe\,Ia is a good candidate for the collisional scenario. Although we determined that two of our spectra were of appropriate quality to evaluate an upper limit for the mass of hydrogen from a nondegenerate companion, our results do not place strong constraints on the progenitor system. We also commented on the redward progression of the peak wavelengths of the iron emission lines at late phases and noted that SN\,2012fr is again the outlier in this category for its apparent blueward shift, but that modeling of nebular ejecta would be necessary to properly interpret this peculiarity. The data released by this publication should also be useful for future conglomerate studies of the growing sample of nebular-phase spectra of SNe\,Ia.

\section*{Acknowledgements}

We thank our anonymous referee for a thoughtful and constructive review.

This research makes use of observations from the Las Cumbres Observatory and is supported by US National Science Foundation (NSF) grant AST--1313484.
We made use of {\it Swift}/UVOT data reduced by P. J. Brown and released in the {\it Swift} Optical/Ultraviolet Supernova Archive (SOUSA). SOUSA is supported by NASA's Astrophysics Data Analysis Program through grant NNX13AF35G.
This work is based in part on observations and discoveries from the the Katzman Automatic Imaging Telescope (KAIT) at Lick Observatory. We are grateful to the staff at Lick Observatory for their assistance. KAIT and its ongoing operation were made possible by donations from Sun Microsystems, Inc., the Hewlett-Packard Company, AutoScope Corporation, Lick Observatory, the NSF, the University of California, the Sylvia \& Jim Katzman Foundation, and the TABASGO Foundation. Research at Lick Observatory is partially supported by a generous gift from Google.

We also used observations obtained at the Gemini Observatory, acquired through the Gemini Observatory Archive through programs GS-2013A-Q-10, GS-2013B-Q-48, GS-2015B-Q-62, and GS-2016A-Q-61 and processed using the Gemini {\sc IRAF} package, which is operated by the Association of Universities for Research in Astronomy, Inc., under a cooperative agreement with the NSF on behalf of the Gemini partnership: the NSF (United States), the National Research Council (Canada), CONICYT (Chile), Ministerio de Ciencia, Tecnolog\'{i}a e Innovaci\'{o}n Productiva (Argentina), and Minist\'{e}rio da Ci\^{e}ncia, Tecnologia e Inova\c{c}\~{a}o (Brazil). We thank the queue service observers and technical support staff at Gemini Observatory for their assistance with the acquisition and reduction of our data.
 
This work is based in part on observations from the Low Resolution Imaging Spectrometer at the Keck-1 telescope. We are grateful to the staff at the Keck Observatory for their assistance, and we extend special thanks to those of Hawaiian ancestry on whose sacred mountain we are privileged to be guests. The W.~M.\ Keck Observatory is operated as a scientific partnership among the California Institute of Technology, the University of California, and NASA; it was made possible by the generous financial support of the W.~M.\ Keck Foundation. We thank Daniel Perley\footnote{http://www.astro.caltech.edu/$\sim$dperley/programs/lpipe.html} and S. Bradley Cenko for the use of, and assistance with, their Keck LRIS imaging and spectroscopy reduction pipelines. We also thank Patrick Kelly, WeiKang Zheng, Isaac Shivvers, and Brad Tucker for their assistance with some of the observations.

The supernova research of A.V.F.'s group at U.C. Berkeley is supported by Gary \& Cynthia Bengier, the Richard \& Rhoda Goldman Fund, the Christopher R. Redlich Fund, the TABASGO Foundation, NSF grant AST--1211916, and the Miller Institute for Basic Research in Science (U.C. Berkeley). 
A.V.F.'s work was conducted in part at the Aspen Center for Physics, 
which is supported by NSF grant PHY-1607611; he thanks the Center for 
its hospitality during the neutron stars workshop in June and July 2017.
This research has made use of the online repository of data from the long-term Berkeley Supernova Project that is publicly available at \url{http://heracles.astro.berkeley.edu/sndb/}.
Support for I.A. was provided by NASA through the Einstein Fellowship Program, grant PF6-170148.
D.J.S. acknowledges support from NSF grant AST--1517649.

This research has made use of the NASA/IPAC Extragalactic Database (NED) which is operated by the Jet Propulsion Laboratory, California Institute of Technology, under contract with NASA.
This work has made use of the Weizmann interactive supernova data repository presented by \cite{2012PASP..124..668Y} and accessible at \url{http://wiserep.weizmann.ac.il}.
This work has made use of the Open Supernova Catalog that is publicly available at \url{https://sne.space/}, presented by \cite{2016arXiv160501054G}. We thank the OSC creators for their quick responses to our questions.
This work has made use of the free, online, collaborative article-writing website \url{www.authorea.com}. We thank the staff at Authorea for their responsive assistance during the preparation of this document.

\bibliographystyle{apj}
\bibliography{apj-jour,myrefs}

\label{lastpage}

\end{document}